\documentclass[lettersize,journal]{IEEEtran}
\usepackage{amsmath,amsfonts}
\usepackage{algorithmic}
\usepackage{algorithm}
\usepackage{array}
\usepackage[caption=false,font=normalsize,labelfont=sf,textfont=sf]{subfig}
\usepackage{textcomp}
\usepackage{stfloats}
\usepackage{url}
\usepackage{verbatim}
\usepackage{graphicx}
\usepackage{cite}
\usepackage{subfiles}
\usepackage{titling}  % if you need separate title pages for the subfiles

\usepackage{tikz}
\usetikzlibrary{
	positioning,
	shapes.geometric,
	arrows.meta,
	calc
}

\usepackage{orcidlink}

\begin{document}

\title{Hybrid epidemic simulation framework coupling equation-based and individual-based models}

\author{
	Jaeyoung Kwak~\orcidlink{0000-0002-9627-3569}, 
	Michael H. Lees~\orcidlink{0000-0002-5457-9180},
	Chin Chun Ooi~\orcidlink{0000-0003-4813-4529}, and
	Wentong Cai~\orcidlink{0000-0002-0183-3835}, \IEEEmembership{Senior Member, IEEE}
	\thanks{Jaeyoung Kwak and Wentong Cai are with College of Computing and Data Science, Nanyang Technological University, Singapore.} % (e-mail: jaeyoung.kwak@ntu.edu.sg; ASWTCAI@ntu.edu.sg)
	\thanks{Michael H. Lees is with Informatics Institute, University of Amsterdam, The Netherlands.} % (e-mail: M.H.Lees@uva.nl)
	\thanks{Chin Chun Ooi is with Institute of High Performance Computing, Agency for Science, Technology and Research, Singapore.} % (e-mail: ooicc@a-star.edu.sg)
}

% \thanks{Manuscript received April 19, 2026; revised August 16, 2021.}}

\markboth{\today}%
{Shell \MakeLowercase{\textit{et al.}}: A Sample Article Using IEEEtran.cls for IEEE Journals}
%\markboth{IEEE Transactions on Computational Social Systems,~Vol.~00, No.~0, Month~YEAR}%

% Remember, if you use this you must call \IEEEpubidadjcol in the second column for its text to clear the IEEEpubid mark.

\maketitle

\begin{abstract} % 100–250 word abstract
Mass gathering events (MGEs) can create intense, short-lived transmission hotspots whose effects propagate through urban populations. Quantifying these effects requires linking fine-scale interactions during gatherings with epidemic transmission across metropolitan mobility networks. We developed a hybrid multiscale simulation framework coupling empirically derived individual-based contact networks for MGEs with a stochastic metapopulation susceptible-exposed-infected-recovered (SEIR) model. Using population and commuting data from Madrid, Spain, we investigated how event-related transmission affects epidemic dynamics and district-level invasion. Across the scenarios examined, MGEs increased epidemic peak incidence and accelerated epidemic progression, although the magnitude of these effects depended on event-specific contact patterns and background transmissibility. Dense contact patterns produced greater epidemic amplification, whereas measures reducing close-contact interactions weakened these effects. MGEs also advanced epidemic invasion across districts and increased the temporal synchronization of arrival times, producing a more synchronized metropolitan epidemic wave. The association between MGEs and earlier district-level invasion was partly explained by the size of the event-generated epidemic seed. The main patterns persisted across the epidemiological parameter ranges examined. The framework provides a multiscale approach for investigating how mass gathering management may influence metropolitan epidemic dynamics.
\end{abstract}


\section{Introduction}
\label{sec:introduction}
Mass gathering events (MGEs), including concerts, sporting events, religious ceremonies, and festivals, represent important settings for infectious disease transmission because they temporarily bring large numbers of people into close proximity~\cite{Johansson_2012,Alshammari_2025}. High crowd density, prolonged contact duration, and intensified social interactions can increase opportunities for transmitting respiratory infections such as influenza, SARS, and COVID-19~\cite{Rhodes_2008,Smieszek_2009,Lipsitch_2009,Rivers_2014,Ng_2021,Crawford_2022,Zheng_2024}. Consequently, restrictions on mass gatherings constituted one of the principal non-pharmaceutical interventions implemented during the COVID-19 pandemic~\cite{McCloskey_2020}. As societies seek to balance social activities with epidemic preparedness, understanding how transient gathering events influence subsequent epidemic dynamics remains an important public-health challenge.

The epidemiological consequences of an MGE, however, extend beyond the duration and physical boundaries of the event itself. Although a gathering may last only a few hours or days, the infections generated during this period can subsequently propagate through daily mobility networks and influence epidemic spread across an entire metropolitan population. Critically, transmission within a gathering is not necessarily well described by homogeneous mixing. Crowd interactions vary over time and space according to movement, activity, venue layout, crowd density, and individual behavior, producing heterogeneous and dynamically evolving contact patterns~\cite{Chowell_2012,Johansson_2012}. Empirical studies of large events have shown, for example, that crowd movement can alternate between periods of movement and rest and that contact-duration distributions can be highly heterogeneous~\cite{Gambaudo_2022}. Similarly, empirical contact measurements combined with individual-based transmission simulations have demonstrated that using observed social-mixing patterns rather than uniform mixing can substantially change simulated transmission outcomes~\cite{Rainey_2021}. Duan~\textit{et.~al.}~\cite{Duan_2024} has also modeled the temporal and spatial structure of human contacts using multilayer agent networks, demonstrating the relevance of structured contact representations for epidemic modeling. These observations indicate that the transmission process within an MGE may require substantially finer temporal and interpersonal resolution than the subsequent metropolitan-scale epidemic.

This difference in required resolution creates a fundamental modeling challenge. Metapopulation compartmental models provide a computationally efficient representation of epidemic transmission across large populations and spatially connected subpopulations. By describing each population unit through aggregate epidemiological compartments, such as susceptible, exposed, infectious, and recovered individuals, they can efficiently capture large-scale transmission and mobility between locations~\cite{Hethcote_2000}. However, the same aggregation that makes these models computationally tractable also limits their ability to represent the detailed contact structure of a specific mass gathering. Individuals within a population patch are typically represented through aggregate mixing rates or population-level transmission parameters, making it difficult to explicitly account for heterogeneous contact duration, time-varying crowd density, activity-dependent interactions, or event-specific contact networks. Thus, a conventional metapopulation model can represent the subsequent propagation of infections through an urban population, but may not adequately resolve the localized transmission process that generates those infections during a gathering.

Individual-based models (IBMs), in contrast, explicitly represent individuals and their interactions, allowing disease transmission to depend on spatial proximity, social-network structure, movement, behavior, and other sources of heterogeneity. For example, agent-based epidemic models have been used to investigate how heterogeneous mobility patterns and individual-level interventions influence disease transmission and outbreak dynamics~\cite{Vedam_2023}. This level of resolution is particularly appropriate for modeling transmission within mass gatherings, where the structure and temporal evolution of interpersonal contacts can strongly influence the number of infections generated during the event~\cite{Chowell_2012,Gambaudo_2022,Rainey_2021}. However, extending such a representation to an entire metropolitan population introduces a different limitation. An urban-scale IBM must track potentially millions of individuals, their epidemiological states, movements, and interactions over many simulation time steps, resulting in substantially greater computational and data requirements than aggregate population models~\cite{Thomine_2021}. Comparative studies of large-scale agent-based and structured metapopulation models highlight a fundamental trade-off: individual-level frameworks capture richer epidemiological detail but incur substantially higher computational demands~\cite{Ajelli_2010}. Consequently, applying an individual-based representation indiscriminately across both the gathering and the surrounding metropolitan population is often impractical.

These complementary limitations motivate a multiscale modeling strategy in which model resolution is allocated selectively according to the spatial and temporal scale of the process being represented. The key methodological question is therefore not whether epidemic transmission should be represented at the individual or population level, but where each level of resolution is necessary. A hybrid approach can represent the short-lived and highly heterogeneous transmission process within a gathering using an individual-based model while describing the subsequent metropolitan epidemic using a computationally efficient metapopulation model. The two representations can then be coupled through the infections generated during the event: the high-resolution event model determines the additional infections produced by the gathering, which become an event-induced epidemic seed for the subsequent population-scale simulation. Such selective-resolution strategies have been proposed more generally to combine the computational efficiency of population-level models with the detailed information available from individual-based representations~\cite{Marilleau_2018,Bicker_2025}. More broadly, this coupling principle provides a general methodology for embedding transient, high-resolution interaction processes within population-scale epidemic models without requiring the entire urban system to be simulated at individual resolution.

Despite these advances, existing epidemic models for MGEs have only partially addressed this multiscale resolution problem. Previous studies have examined the epidemiological consequences of mass gatherings using compartmental, metapopulation, network, and individual-based approaches, but the detailed contact heterogeneity within an event and its subsequent propagation through metropolitan mobility networks are often represented separately or at different levels of abstraction~\cite{Chowell_2012,Lim_2021}. Consequently, the mechanistic pathway linking event-specific contact structure, infections generated during the gathering, and subsequent metropolitan epidemic progression remains insufficiently characterized. In particular, relatively little attention has been paid to treating the infections generated during a gathering as an \emph{event-induced epidemic seed} and quantitatively evaluating how variation in this seed influences the timing and spatial progression of epidemic invasion across the receiving urban population. A framework that explicitly couples these scales is therefore needed to connect transient local transmission processes with their longer-term metropolitan consequences.

To address this gap, we develop a multiscale hybrid simulation framework that spatially and temporally couples an individual-based model of MGE transmission with a stochastic metapopulation susceptible-exposed-infected-recovered (SEIR) model of urban epidemic spread. Individual-level contact networks derived from empirical MGE datasets are used to represent transmission during gathering events, thereby preserving event-specific heterogeneity in interpersonal interactions. The infections generated during the event are then transferred to the metropolitan-scale model as an event-induced epidemic seed, after which epidemic transmission proceeds through the urban population and commuter mobility network. This coupling allows the framework to resolve transmission at the scale at which contact heterogeneity is most consequential while retaining computational tractability for subsequent city-wide epidemic simulation.

We demonstrate the framework using Madrid, Spain, by combining real-world population distributions and commuter flows with empirically derived contact networks representing several types of MGEs. We first examine how different gathering scenarios modify epidemic seeding, epidemic magnitude, temporal progression, and spatial spread. We then investigate district-level epidemic invasion timing using survival analysis to quantify how mass gathering events are associated with earlier and more synchronized epidemic invasion, and to examine the extent to which this association is related to the size of the event-generated epidemic seed and the subsequent transmission environment. Finally, we assess the robustness of these relationships across a range of background and event-specific transmission conditions.

The contributions of this study are threefold. First, we develop a multiscale modeling framework that couples high-resolution individual-based representation of transient gathering events with a computationally tractable metropolitan-scale epidemic model. Second, we quantify how heterogeneous contact structures associated with different gathering scenarios influence event-generated epidemic seeding, epidemic magnitude, temporal progression, and spatial invasion. Third, by combining epidemic simulation with survival analysis, we provide a quantitative interpretation of the relationship between event-generated epidemic seeding and subsequent metropolitan epidemic invasion, demonstrating how localized transmission processes can produce measurable city-wide consequences.

\section{Materials and Methods}
\label{sec:methods}

\subsection{Hybrid simulation framework}
\label{sec:methods_overview}

\begin{figure*}
\centering
%\resizebox{\textwidth}{!}{%
\resizebox{17cm}{!}{%
\begin{tikzpicture}[
node distance=1.5cm and 2cm,
auto,
block/.style={
rectangle,
draw,
text width=4cm,
align=center,
rounded corners,
minimum height=4em,
font=\sffamily\large
},
decision/.style={
diamond,
draw,
text width=3cm,
align=center,
aspect=2.5,
font=\sffamily\large
},
event_block/.style={
rectangle,
draw,
text width=4cm,
align=center,
rounded corners,
minimum height=4em,
font=\sffamily\large
},
line/.style={
-Stealth,
thick,
rounded corners
},
label_node/.style={
font=\sffamily\large\bfseries,
fill=white,
inner sep=2pt
}
]

% Nodes: left-to-right flow
\node[block] (start) {Start Step $t$};

\node[decision, right=of start] (decide)
{Event Day?};

% No-event branch
\node[block, above right=1cm and 3cm of decide]
(std_model) {Metapopulation\\SEIR Dynamics};

% Event-day branch
\node[block, below right=1cm and 1cm of decide]
(split) {Recruit / Import\\and split\\(Attendees /\\Non-attendees)};

% Parallel processes
\node[block, right=2cm of split, yshift=1.8cm]
(non_att) {Metapopulation\\SEIR Dynamics\\(Non-attendees)};

\node[event_block, right=2cm of split, yshift=-1.8cm]
(att) {Individual-based model\\(Attendees)};

% Merge and removal
\node[block, right=2cm of non_att, yshift=-1.8cm]
(merge) {Reintegrate\\(Resident Attendees)};

\node[block, below=of merge, text width=3.5cm]
(remove) {Remove Imported\\Attendees};

% End node
\node[block] (end)
at ($(merge.east |- start) + (2.5cm,0)$)
{Next Step $t + \Delta t$};

% Main connections
\draw[line] (start) -- (decide);

\draw[line] (decide) |-
node[label_node, pos=0.2, above=0.1cm] {No}
(std_model);

\draw[line] (std_model) -| (end);

\draw[line] (decide) |-
node[label_node, pos=0.2, below=0.1cm] {Yes}
(split);

% Connections to parallel processes
\draw[line] (split.east) -- ++(0.5,0) |- (non_att.west);
\draw[line] (split.east) -- ++(0.5,0) |- (att.west);

% Non-attendee process to merge
\draw[line] (non_att.east) -- ++(0.5,0) |- (merge.west);

% Branch after the individual-based model
\coordinate (branch_point) at ($(att.east) + (0.5cm,0)$);
\draw[line] (att.east) -- (branch_point);

% Resident attendees
\draw[line] (branch_point) |-
node[label_node, pos=0.0,
xshift=-0.1cm, yshift=1.1cm] {Residents}
(merge.west);

% Imported attendees
\draw[line] (branch_point) |-
node[label_node, pos=0.0,
xshift=-1.0cm, yshift=-1.5cm] {Imported}
(remove.west);

% Merge to end
\draw[line] (merge.east) -- ++(2.5cm,0) -- (end.south);

\end{tikzpicture}%
}

\caption{Schematic representation of the hybrid simulation logic. On event days, a subset of the population is recruited into an individual-based transmission model while the remaining population continues metapopulation susceptible-exposed-infected-recovered (SEIR) dynamics. Updated epidemiological states are reintegrated before advancing to the next time step.}
\label{fig:hybrid_model_overview}
\end{figure*}

The framework couples a city-scale stochastic metapopulation susceptible--exposed--infectious--recovered (SEIR) model with an individual-based transmission model for mass gathering events (MGEs), using a split--apply--merge workflow (Fig.~\ref{fig:hybrid_model_overview}). The metapopulation model represents Madrid as a network of 21 administrative districts connected by daily commuter flows. On non-event days, the entire population follows the metapopulation SEIR dynamics. On event days, resident attendees are temporarily removed from their home--destination subpopulations and simulated together with imported infectious attendees using an individual-based contact network. Non-attendees continue to follow the metapopulation dynamics during the same time step.

At the end of the event, resident attendees are returned to their home districts with their updated epidemiological states. They are then redistributed across destination districts according to the mobility matrix. Imported attendees are removed from the modeled population and do not contribute directly to subsequent community transmission. Newly exposed resident attendees, however, remain in the metapopulation model and can transmit infection during subsequent simulation steps. This coupling enables event-level transmission to alter the initial conditions for subsequent city-scale epidemic spread.

% \subsection{Population, mobility, and event data}
\subsection{Dataset}
\label{sec:dataset}

The metapopulation model uses population and commuter-flow data for the 21 administrative districts of Madrid~\cite{MiguelArribas_2023}. Let $N_i$ denote the resident population of district $i$, and let $D_{ij}$ denote the probability that a resident of district $i$ travels to district $j$ during the away-from-home phase. The population and mobility data are provided in the Supplementary Materials.

Event-level contact patterns were derived from empirical pedestrian trajectory datasets. A contact was recorded when two attendees were within 1.5\,m of each other, and the cumulative duration of such contacts was used to weight the edges of the event contact network. We considered five scenarios: an Amsterdam football match ($N_{\mathrm{event}}=362$), an Amsterdam dance event ($N_{\mathrm{event}}=1048$)~\cite{Rutten_2022}, and three indoor concert scenarios from Leipzig ($N_{\mathrm{event}}=1194$, 1158, and 1054)~\cite{Moritz_2021}. The Leipzig scenarios represent progressively stronger contact-mitigation measures, from unrestricted movement and fully occupied seating to restricted movement and increased distancing. The full event schedules and contact-network summary statistics are reported in the Supplementary Materials.

\subsection{Stochastic metapopulation SEIR model}
\label{sec:metapopulation_model}

The population is disaggregated into home--destination subpopulations $N_{ij}$, where $i$ denotes the home district and $j$ the destination district during the away-from-home phase. The model uses a time step of $\Delta t=0.5$ days, comprising an away phase, during which individuals mix in their destination districts, and a home phase, during which they return to their home districts. Individuals in each subpopulation occupy one of four epidemiological states: susceptible ($S_{ij}$), exposed ($E_{ij}$), infectious ($I_{ij}$), or recovered ($R_{ij}$).

Transmission is governed by the phase-specific force of infection
\begin{equation}
	\lambda_{\ell}^{p}(t)
	=
	\beta\frac{I_{\ell}^{p}(t)}{N_{\ell}^{p}(t)},
	\qquad p\in\{A,H\},
	\label{eq:infection-force-general}
\end{equation}
where $\beta$ is the community transmission rate, and $I_{\ell}^{p}(t)$ and $N_{\ell}^{p}(t)$ denote the infectious and total populations present in district $\ell$ during phase $p$. During the away phase, populations are aggregated by destination district; during the home phase, they are aggregated by home district. Thus, the force of infection experienced by subpopulation $(i,j)$ is $\lambda_j^A(t)$ during the away phase and $\lambda_i^H(t)$ during the home phase.

Epidemiological transitions are stochastic. During each half-day phase, the number of susceptible individuals becoming exposed is drawn from a binomial distribution:
\begin{align}
	p_{ij}^{E}(t)
	&=1-\exp[-\lambda_{ij}(t)\Delta t],\\
	\Delta E_{ij}(t)
	&\sim \mathrm{Binomial}\bigl(S_{ij}(t),p_{ij}^{E}(t)\bigr),
\end{align}
where $\lambda_{ij}(t)$ is the corresponding phase-specific force of infection. Exposed-to-infectious and infectious-to-recovered transitions are modeled analogously, with transition probabilities $1-\exp(-\alpha\Delta t)$ and $1-\exp(-\gamma\Delta t)$, respectively. Here, $\alpha=T_{\mathrm{inc}}^{-1}$ and $\gamma=T_{\mathrm{inf}}^{-1}$ are determined by the mean incubation period and infectious period. The simulation uses mean latent and infectious periods of four and five days, respectively. Further details of the transition equations are provided in the Supplementary Materials.

For the survival analysis, we also calculate daily transmission pressure in district $i$ by averaging the away- and home-phase forces of infection:
\begin{equation}
	\Lambda_i(k)=\frac{1}{2}
	\left[\lambda_i^A(k)+\lambda_i^H(k)\right],
	\label{eq:daily-infection-force}
\end{equation}
where $k$ denotes calendar day. This measure summarizes the simulated transmission environment at daily resolution and is used to construct a pre-invasion covariate in the Cox models described below.

\subsection{Individual-based event transmission and hybrid coupling}
\label{sec:ibm}
\label{sec:coupling}

Each event is represented as a weighted contact network. For a susceptible attendee $j$, cumulative contact duration with infectious attendees is
\begin{equation}
	\tau_j=\sum_{a\in\mathcal{I}_{\mathrm{event}}}\tau_{aj},
\end{equation}
where $\mathcal{I}_{\mathrm{event}}$ is the set of infectious attendees and $\tau_{aj}$ is the cumulative duration of close contact between attendees $a$ and $j$. Assuming transmission follows a Poisson process, the probability that attendee $j$ becomes exposed during the event is
\begin{equation}
	P_{j,\mathrm{inf}}=1-\exp(-\beta_{\mathrm{event}}\tau_j),
	\label{eq:event-transmission}
\end{equation}
where $\beta_{\mathrm{event}}$ is the event-specific transmission rate. A Bernoulli trial determines whether each susceptible attendee becomes exposed; updated epidemiological states are then used when resident attendees are returned to the metapopulation.

On each event day, resident attendees are sampled from the Madrid population. The number recruited from each home district is determined using a gravity-based probability,
\begin{equation}
	p_i=
	\frac{N_i/d_{iv}^{\eta}}
	{\sum_k N_k/d_{kv}^{\eta}},
	\label{eq:gravity-recruitment}
\end{equation}
where $d_{iv}$ is the distance from district $i$ to the event venue in district $v$, and the distance-decay exponent is $\eta=1.5$. The number recruited from each district is set to $n_i=\mathrm{round}(p_iN_{\mathrm{event}}^{\mathrm{res}})$. Attendees are sampled from the district's SEIR compartments in proportion to their current sizes.

In addition to resident attendees, $I_{\mathrm{ss}}$ infectious individuals are introduced as imported attendees. They participate in event transmission but are not assigned to a Madrid district. Resident attendees are temporarily removed from the metapopulation state arrays while the event is simulated. After the event, they are returned to their home districts and redistributed across destination districts using the mobility probabilities $D_{ij}$. Imported attendees are removed from the system after the event. Consequently, imported infections can generate secondary infections among resident attendees, which can then propagate through subsequent community transmission.

\subsection{Statistical analysis of epidemic arrival times}
\label{sec:statistical-analysis}

For each stochastic simulation replicate, district-level epidemic arrival time, $T_i^{\mathrm{arr}}$, was defined as the first time district $i$ contained at least one infectious individual ($I_i\geq1$). Districts without an infectious individual by the end of the simulation were treated as right-censored. Kaplan--Meier estimators were used to compare the distribution of arrival times across the baseline and MGE scenarios. Differences between survival curves were assessed using a global log-rank test and pairwise log-rank tests with Holm correction.

We then fitted Cox proportional hazards models to quantify the association between MGEs and epidemic invasion timing:
\begin{equation}
	h_i(t)=h_0(t)\exp\left(\mathbf{x}_i^\top\boldsymbol{\theta}\right),
	\label{eq:cox-general}
\end{equation}
where $h_i(t)$ is the hazard of epidemic invasion, $h_0(t)$ is the baseline hazard, and $\mathbf{x}_i$ contains the explanatory variables. A hazard ratio (HR) above one indicates an association with earlier invasion. Because districts within a simulation replicate share the same epidemic realization and transmission history, all Cox models used cluster-robust standard errors with the simulation replicate as the clustering unit~\cite{Lin_1989}.

Three sequential models were fitted. Model M0 included only the MGE indicator, $G_i$. Model M1 additionally included the logarithm of the event-generated epidemic seed, $\log(Z_i+1)$, where $Z_i$ is the total number of exposed and infectious individuals generated directly by the event. Model M2 further included the mean daily transmission pressure during the $W$ days preceding district invasion:
\begin{equation}
	\bar{\Lambda}_i^{(W)}
	=\frac{1}{W}\sum_{k=1}^{W}
	\Lambda_i(T_i^{\mathrm{arr}}-k),
	\label{eq:avg-infection-force}
\end{equation}
where $W=3$, 7, or 14 days. These sequential models were designed to examine how the estimated MGE association changed after adjustment for event-generated seed size and whether recent transmission pressure provided additional statistical information. Because the pressure covariate is defined retrospectively relative to observed arrival time, M2 was treated as an exploratory analysis rather than a prospective prediction model or a causal analysis.

Model fit and discrimination were assessed using the Akaike Information Criterion (AIC) and Harrell's concordance index (C-index), respectively. M1 was designated the principal adjusted model for interpreting the association between MGEs and invasion timing, while M2 was used to assess the additional information provided by recent transmission pressure.

To examine sensitivity to epidemiological conditions, we repeated the survival analysis for the AMS\_dance scenario across 25 combinations of $R_0=1.0$--3.0, in increments of 0.5, and $\beta_{\mathrm{event}}=0.1$--0.5, in increments of 0.1. The resulting parameter-only model (P0) and seed-adjusted model (P1) were compared to assess how adjustment for event-generated seed size changed the association between transmission conditions and arrival timing. We also repeated the principal Cox analysis for three seed districts with contrasting spatial and network characteristics. Detailed model specifications and the complete sensitivity and seed-location results are reported in the Supplementary Materials.

All survival analyses were conducted in Python using the \texttt{lifelines} package.

\subsection{Data and code availability}
\label{data-and-code}
All datasets analyzed in this study are publicly available. The Madrid commuter-flow dataset is available at~\url{https://github.com/phononautomata/madrid}, the Amsterdam event dataset at~\url{https://github.com/philiprn/Crowd-Epidemiology}, and the Leipzig experiment dataset at~\url{https://zenodo.org/records/5137667}. The code for data processing, simulation, and analysis is available at~\url{https://github.com/jaeyoung82/hybrid-epidemic-sim} and archived at Zenodo:~\url{https://doi.org/10.5281/zenodo.22135798}.

%=========================================================
% RESULTS
%=========================================================
\section{Results}
\label{sec:results}
Unless otherwise stated, simulations began with a fully susceptible population and $I_{\mathrm{ss}}=1$ infectious individual imported from outside Madrid. Results are reported for $R_0=1.5$ and $\beta_{\mathrm{event}}=0.5$, based on 1000 independent stochastic simulations per scenario.

\subsection{Event-generated epidemic seeding and temporal progression}
\label{sec:results-spatiotemporal}

\begin{figure}
	\centering
	\includegraphics[width=\columnwidth] {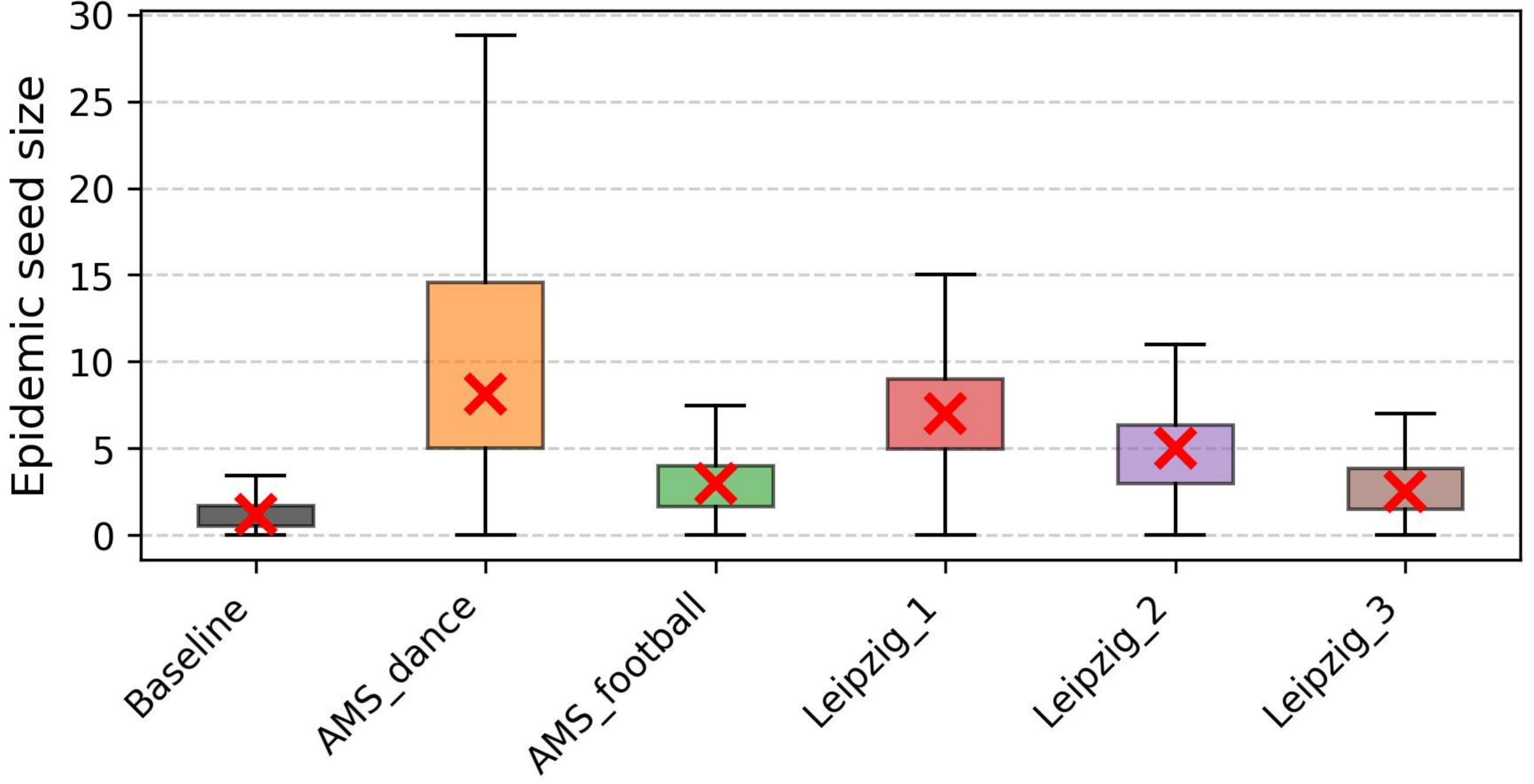} 
	\caption{Distribution of event-generated epidemic seeds immediately after each mass gathering event (MGE), measured as exposed and infectious individuals in Madrid. Boxes indicate the interquartile range (IQR), crosses the median, and whiskers extend to 1.5 IQR. Results are based on 1000 simulations with $R_0=1.5$, $\beta_{\mathrm{event}}=0.5$, and $I_{\mathrm{ss}}=1$.}
	\label{fig:MGE-seeding}
	\vspace{-0.2cm}
\end{figure}

We first examined how different mass gathering events (MGEs) were associated with epidemic seeding and subsequent temporal progression at the metropolitan scale. Fig.~\ref{fig:MGE-seeding} summarizes the number of exposed and infectious individuals generated immediately after each event. All MGE scenarios produced larger epidemic seeds than the baseline scenario without a gathering event, although the seed size varied considerably across event types. AMS\_dance generated the largest seed, followed by Leipzig\_1, Leipzig\_2, Leipzig\_3, and AMS\_football. These differences are consistent with variation in event-specific contact structures and provide different initial conditions for subsequent epidemic evolution.

\begin{figure}
	\begin{minipage}[t]{\columnwidth}
		(a)\\
		\vspace{0.1cm}
		\includegraphics[width=\linewidth]{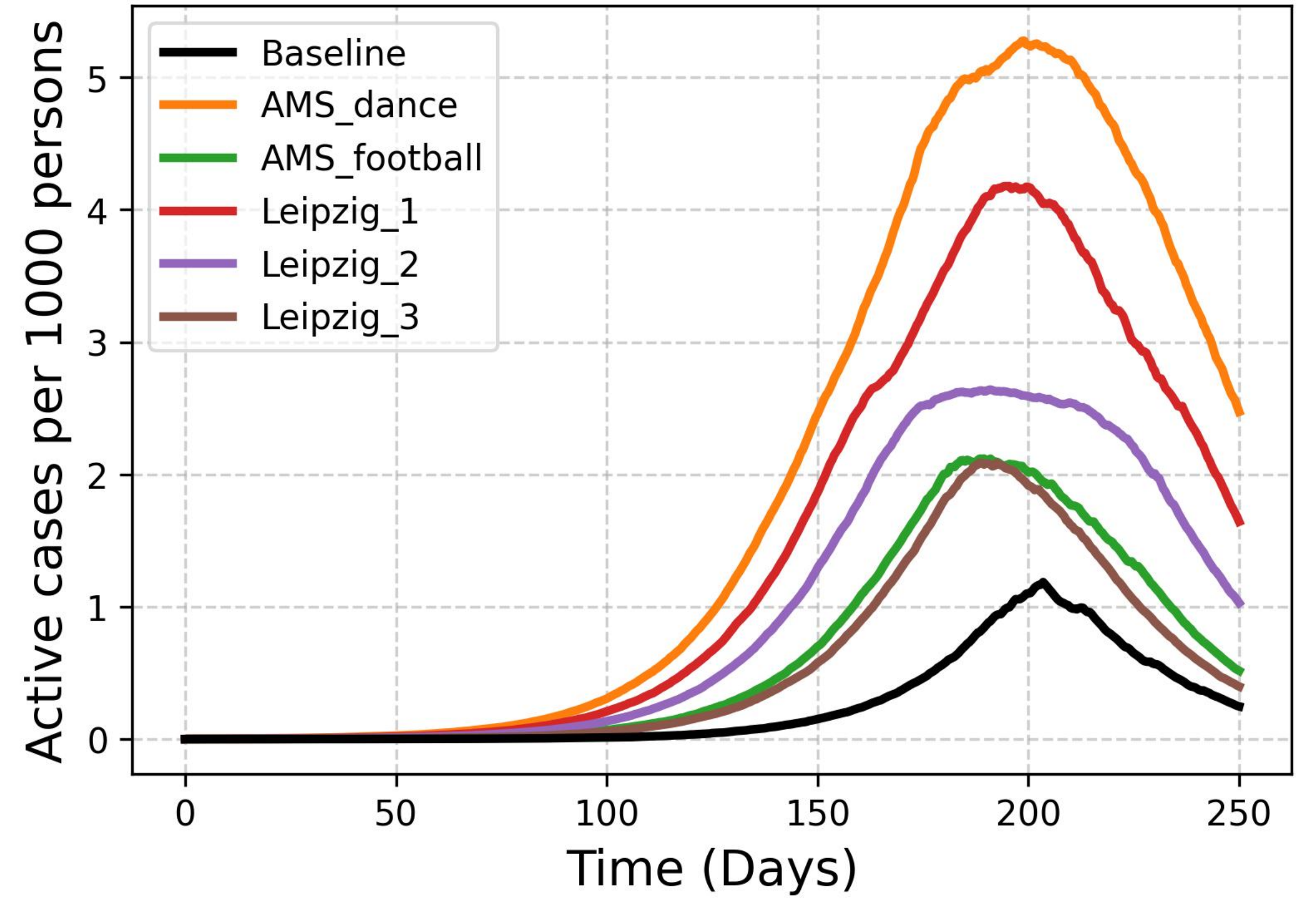}
	\end{minipage}
	\\
	\begin{minipage}[t]{\columnwidth}
		(b)\\
		\vspace{0.1cm}	
		\includegraphics[width=\linewidth]{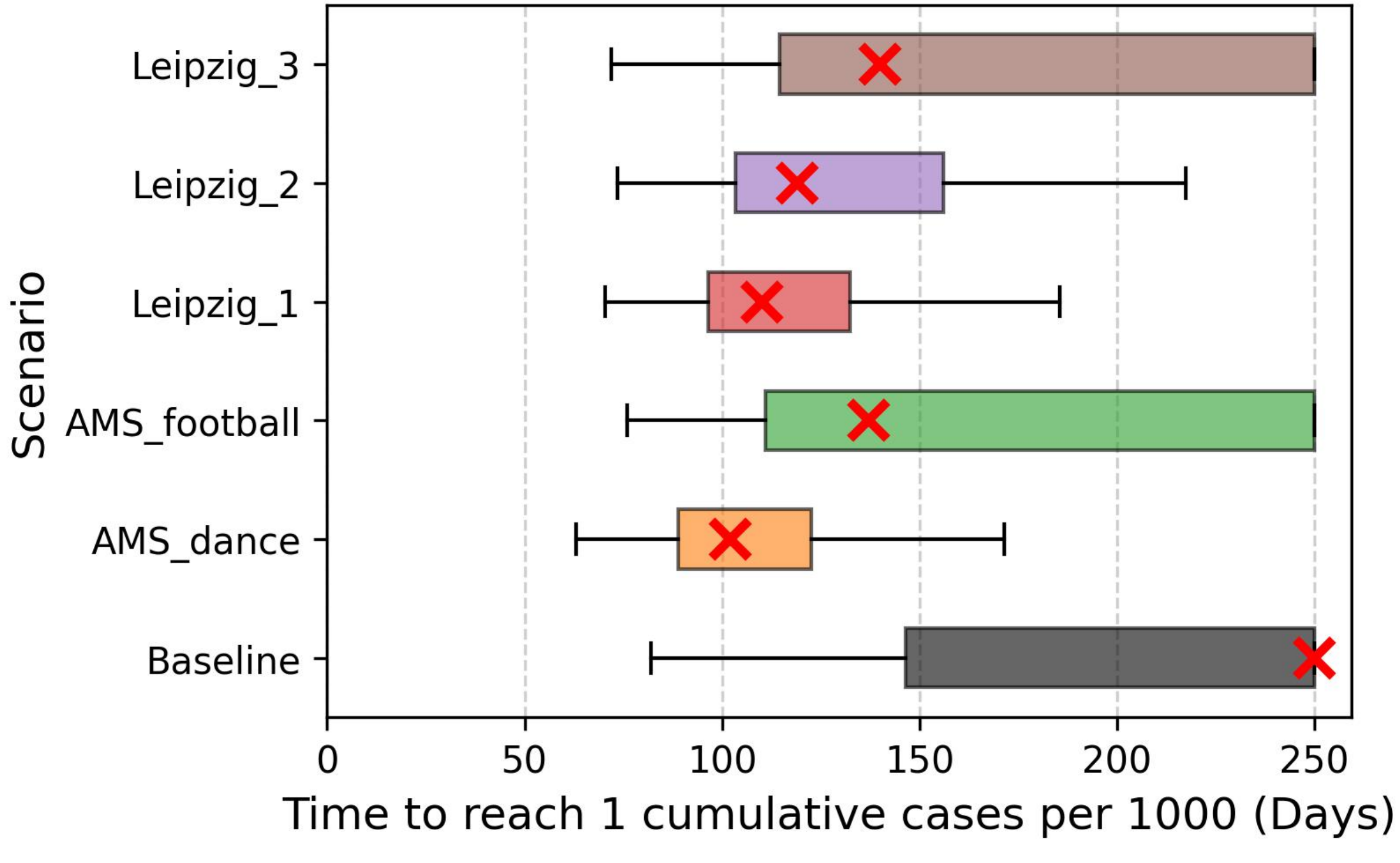}
	\end{minipage}
	\caption{Temporal epidemic dynamics under different MGE scenarios: (a) time evolution of active cases (exposed and infectious individuals) per 1000 residents, comparing trajectories derived from various MGE scenarios against the baseline scenario. (b) time required to reach the epidemic threshold of one cumulative case per 1000 residents. Boxplots show the IQR, median (crosses), and whiskers extending to 1.5 IQR. Mean active-case trajectories include simulations in which the epidemic became established. Results are based on 1000 simulations with $R_0=1.5$, $\beta_{\mathrm{event}}=0.5$, and $I_{\mathrm{ss}}=1$.}
	\label{fig:temporal-progression}
	\vspace{-0.2cm}
\end{figure}

The large epidemic seeds produced by MGEs resulted in substantial changes in the temporal evolution of epidemics. Fig.~\ref{fig:temporal-progression}(a) compares the temporal trajectories of active cases, defined as exposed and infectious individuals, under the MGE and baseline scenarios. All MGE scenarios exhibited faster epidemic growth and higher epidemic peaks than the baseline scenario. The magnitude and timing of the peaks varied across scenarios, with AMS\_dance exhibiting the largest peak incidence. Following Ref.~\cite{Mazzoli_2021}, mean active-case trajectories were calculated using simulations in which the epidemic became established; they therefore describe established outbreaks rather than all simulation runs.

To evaluate how control measures modulate epidemic timing, Fig.~\ref{fig:temporal-progression}(b) shows the time required to reach the threshold of one cumulative case per 1000 residents. Among simulations that crossed this threshold, the MGE scenarios generally reached it earlier than the baseline scenario. Within the Leipzig scenarios, progressively stronger contact-mitigation measures were associated with longer times to threshold, consistent with slower epidemic progression under stronger mitigation. The proportion of simulation runs exceeding the threshold was also higher in the MGE scenarios (60--90\%) than in the baseline scenario (approximately 32\%). These results show that the simulated MGEs were associated with larger event-generated epidemic seeds, faster temporal progression, and a greater likelihood of exceeding the epidemic threshold. Additional results of temporal dynamcis are provided in the Supplementary Materials.

\subsection{Spatial patterns of epidemic invasion}
\label{sec:spatial-invasion}

To examine the spatial progression of epidemics, we compared district-level epidemic arrival times across the baseline and five representative MGE scenarios. Epidemic arrival time was defined as the first simulation time at which a district contained at least one infectious individual. We examined the spatial distribution of mean arrival times, differences relative to the baseline, and the temporal organization of invasion across districts.

\begin{figure*}
	\centering
	\includegraphics[width=0.8\textwidth]{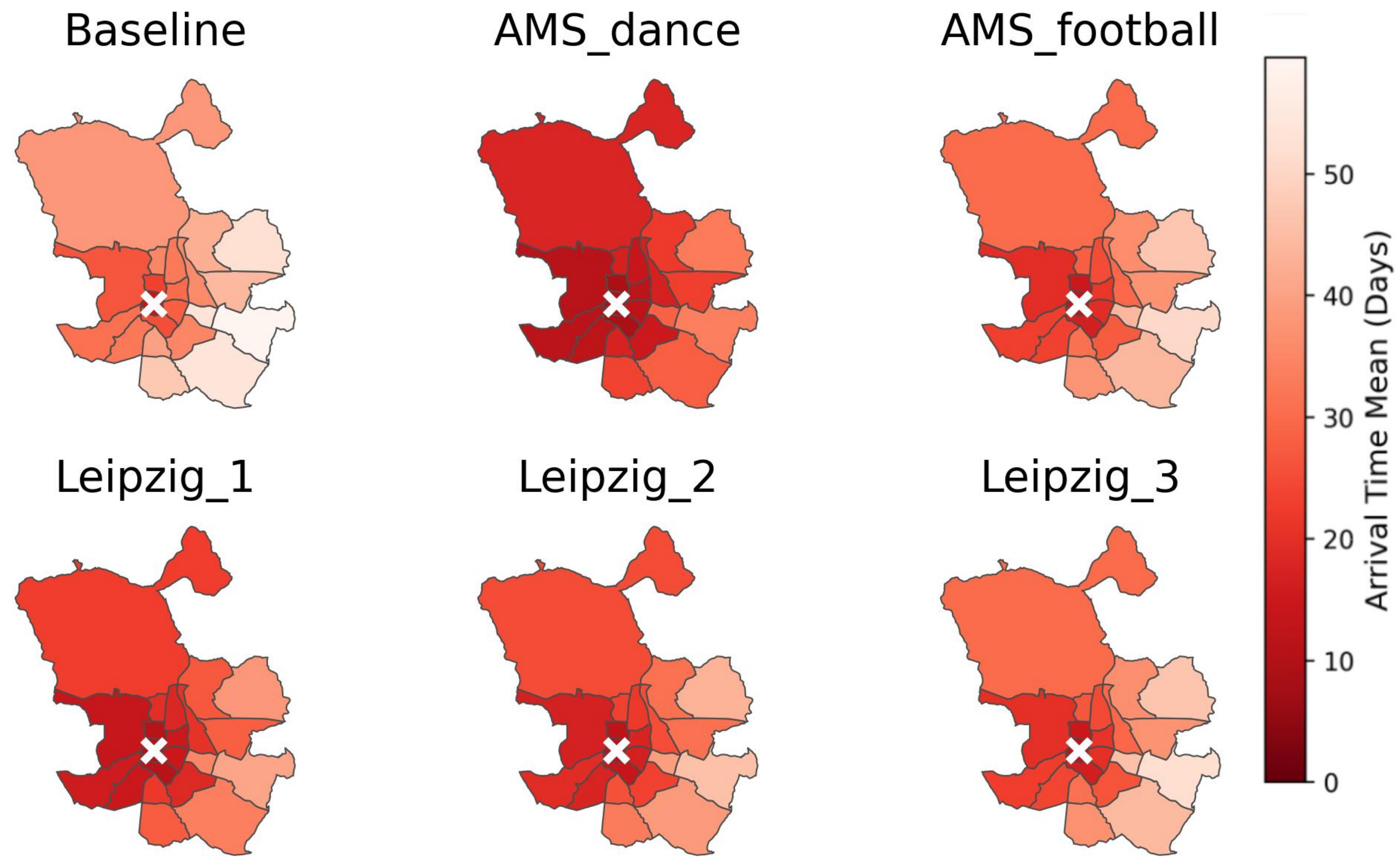}
	\caption{Spatial distribution of mean district-level epidemic arrival time. Darker red indicates earlier arrival. Results are averaged over 1000 simulations with $R_0=1.5$, $\beta_{\mathrm{event}}=0.5$, and $I_{\mathrm{ss}}=1$.}
	\label{fig:arrival-time}
	\vspace{-0.2cm}
\end{figure*}

Fig.~\ref{fig:arrival-time} shows the spatial distribution of mean district-level epidemic arrival times across the Madrid metropolitan area. Under the baseline scenario, the mean arrival times indicate a gradual progression from central districts toward peripheral areas. All MGE scenarios showed earlier mean arrival times across much of the metropolitan area, although the magnitude of the shift varied across scenarios. AMS\_dance produced the largest advances in most districts, particularly in districts reached relatively late under the baseline. In contrast, the spatial arrival-time pattern under Leipzig\_3, which represents the strongest contact-mitigation scenario, was closer to the baseline pattern.

\begin{figure}
\begin{minipage}[t]{\columnwidth}
(a)\\
\includegraphics[width=\linewidth] {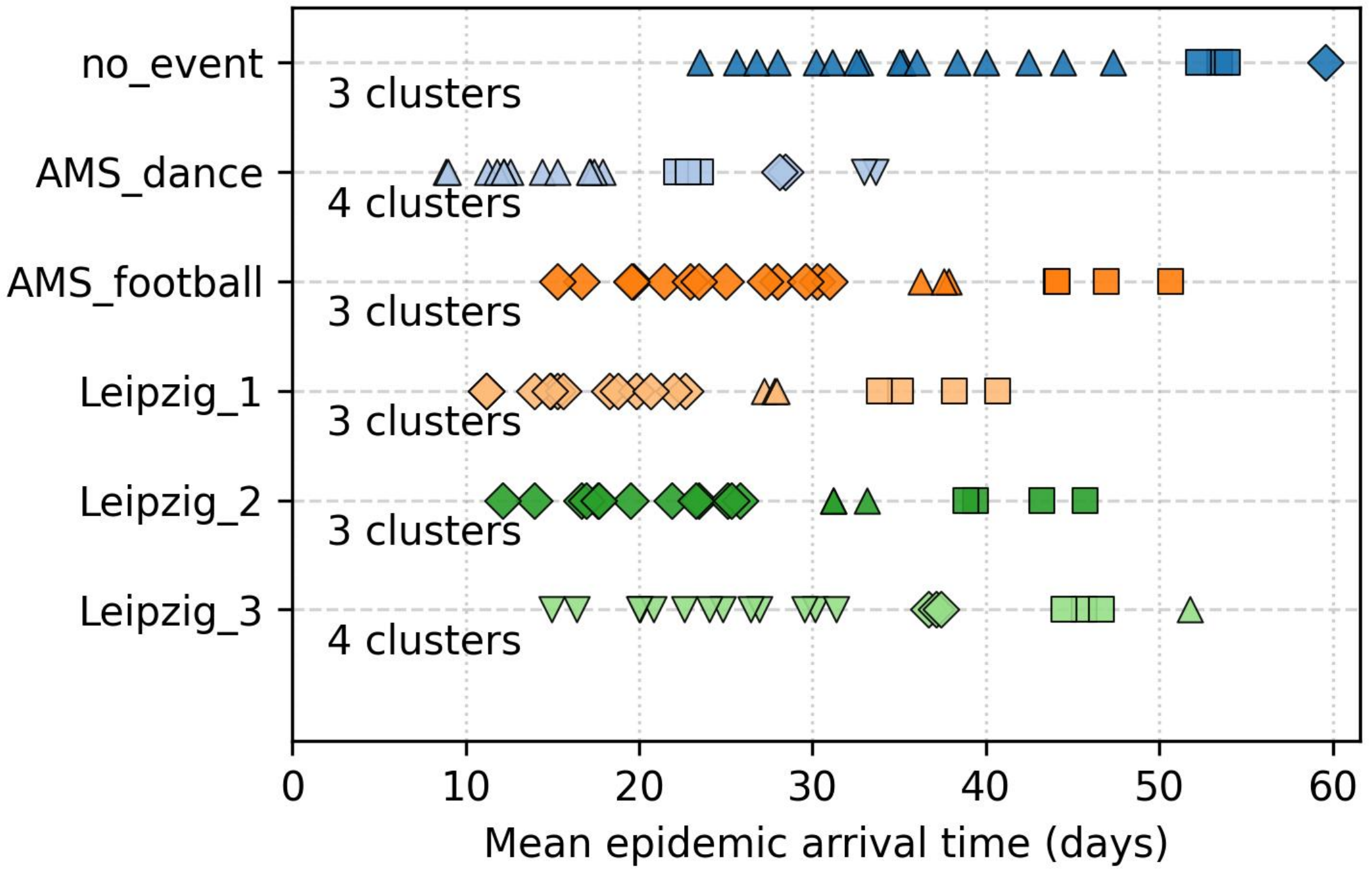}
\end{minipage}
\\
% \hfill
\begin{minipage}[t]{\columnwidth}
(b)\\
\includegraphics[width=\linewidth] {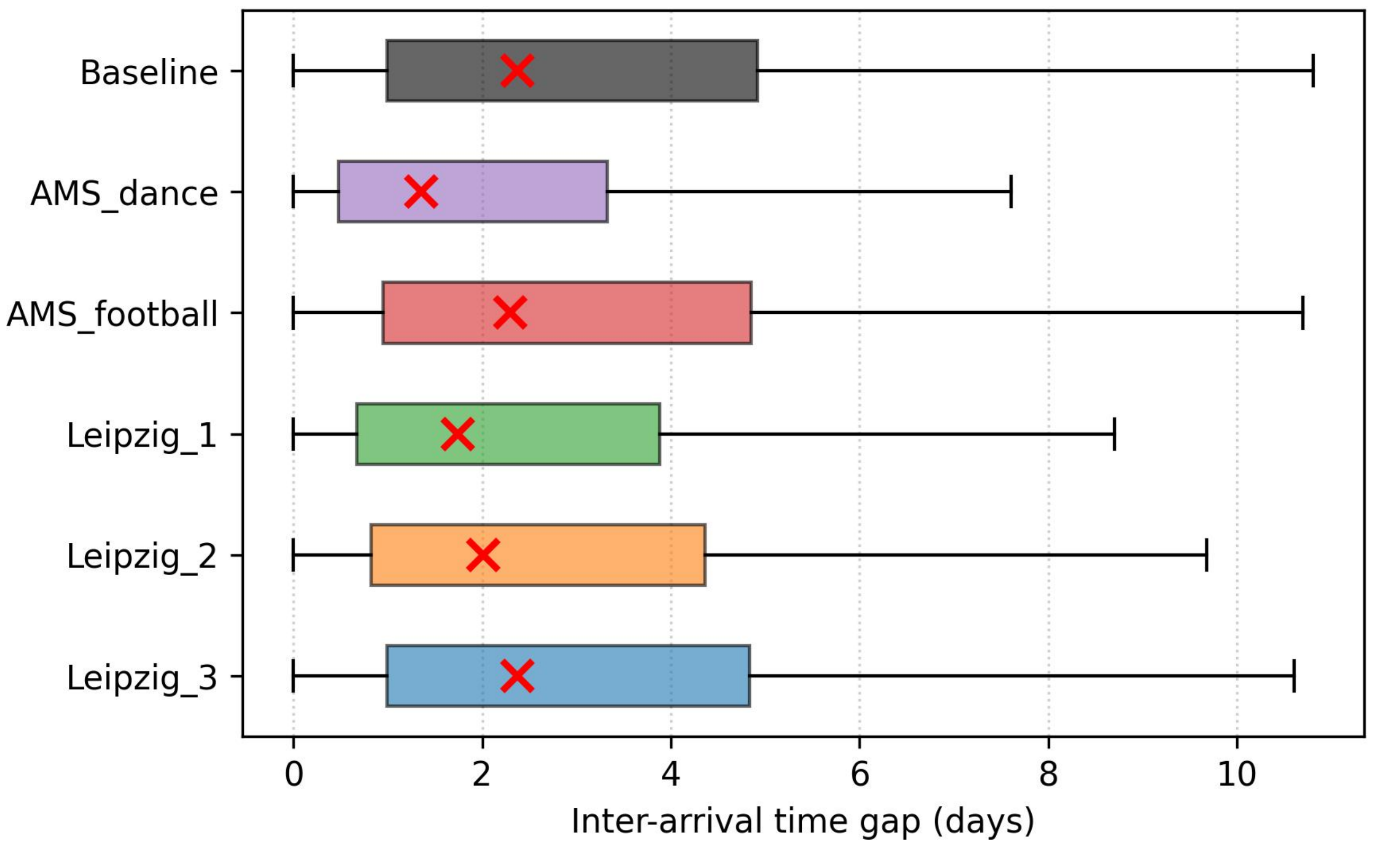}
\end{minipage}
\caption{District-level arrival-time patterns under different MGE scenarios. (a) Mean arrival times for individual districts, grouped using a four-day arrival-time threshold; colors indicate clusters. (b) Distribution of intervals between mean district-level arrival times. Boxes indicate the IQR, crosses the median, and whiskers extend to 1.5 IQR. Results are based on 1000 simulations with $R_0=1.5$, $\beta_{\mathrm{event}}=0.5$, and $I_{\mathrm{ss}}=1$.}
\label{fig:epidemic-arrival-distribution}
\end{figure}

To characterize the temporal organization of district-level invasion, we grouped districts according to their mean arrival times using a four-day threshold, approximately corresponding to one epidemic generation in the simulation model. As shown in Fig.~\ref{fig:epidemic-arrival-distribution}(a), the grouping patterns differed between the baseline and MGE scenarios, with the MGE scenarios generally exhibiting greater temporal clustering of district arrivals. This pattern was most pronounced under AMS\_dance and AMS\_football, whereas stronger contact mitigation across the Leipzig scenarios was associated with less pronounced synchronization. The Leipzig\_3 pattern was comparatively similar to the baseline, suggesting that stronger mitigation can limit the temporal compression of spatial epidemic progression. Fig.~\ref{fig:epidemic-arrival-distribution}(b) complements this by summarizing the intervals between mean district-level arrival times. All MGE scenarios exhibited shorter median intervals than the baseline scenario, indicating a more compressed sequence of mean district arrivals. AMS\_dance showed the shortest interval, whereas AMS\_football and Leipzig\_3 retained relatively longer intervals. These results indicate that the timing and synchronization of spatial invasion varied across event scenarios, with stronger contact mitigation generally associated with a pattern closer to the baseline.

\begin{figure*}[!t]
\centering
\includegraphics[width=0.85\textwidth]{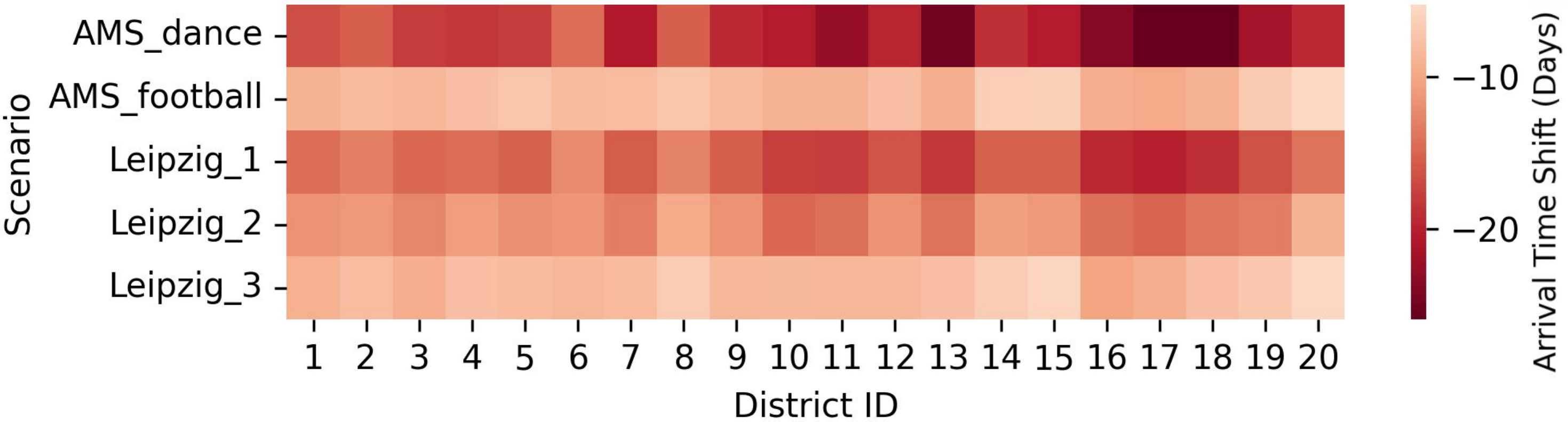}
\caption{Changes in mean district-level arrival time under each MGE scenario relative to the no-event baseline. Darker blue indicates earlier arrival. Results are averaged over 1000 simulations with $R_0=1.5$, $\beta_{\mathrm{event}}=0.5$, and $I_{\mathrm{ss}}=1$.}
\label{fig:arrival-diff-heatmap}
\end{figure*}

Fig.~\ref{fig:arrival-diff-heatmap} quantifies the shifts in mean district-level arrival time relative to the baseline. The largest advances were again observed under AMS\_dance, particularly in districts with later mean arrival times under baseline conditions. By comparison, Leipzig\_3 produced smaller shifts across most districts. Thus, the magnitude of the spatial arrival-time shift varied with the MGE scenario and the district-specific pattern of epidemic progression. A supplementary analysis of AMS\_dance further examined how these shifts varied with the initial seed location; details are provided in the Supplementary Materials.

\subsection{Analysis of epidemic arrival times}
\label{sec:results-arrival-timing-analysis}
To further investigate how MGEs influence the spatial progression of epidemics, we analyzed district-level epidemic arrival times using Kaplan--Meier estimators and sequential Cox proportional hazards models. The Kaplan--Meier analysis compares the timing of epidemic invasion across MGE scenarios, while the Cox models quantify the association between MGE presence and invasion timing under progressively richer specifications. Model M0 estimates the unadjusted association between MGE presence and invasion timing, Model M1 additionally adjusts for the logarithm of the event-generated epidemic seed size, and Model M2 further incorporates mean daily transmission pressure over a specified pre-invasion window. All Cox models used Huber--White sandwich standard errors clustered by simulation run to account for dependence among districts within the same simulation~\cite{Lin_1989}.

\begin{figure}
\centering
\includegraphics[width=\columnwidth]{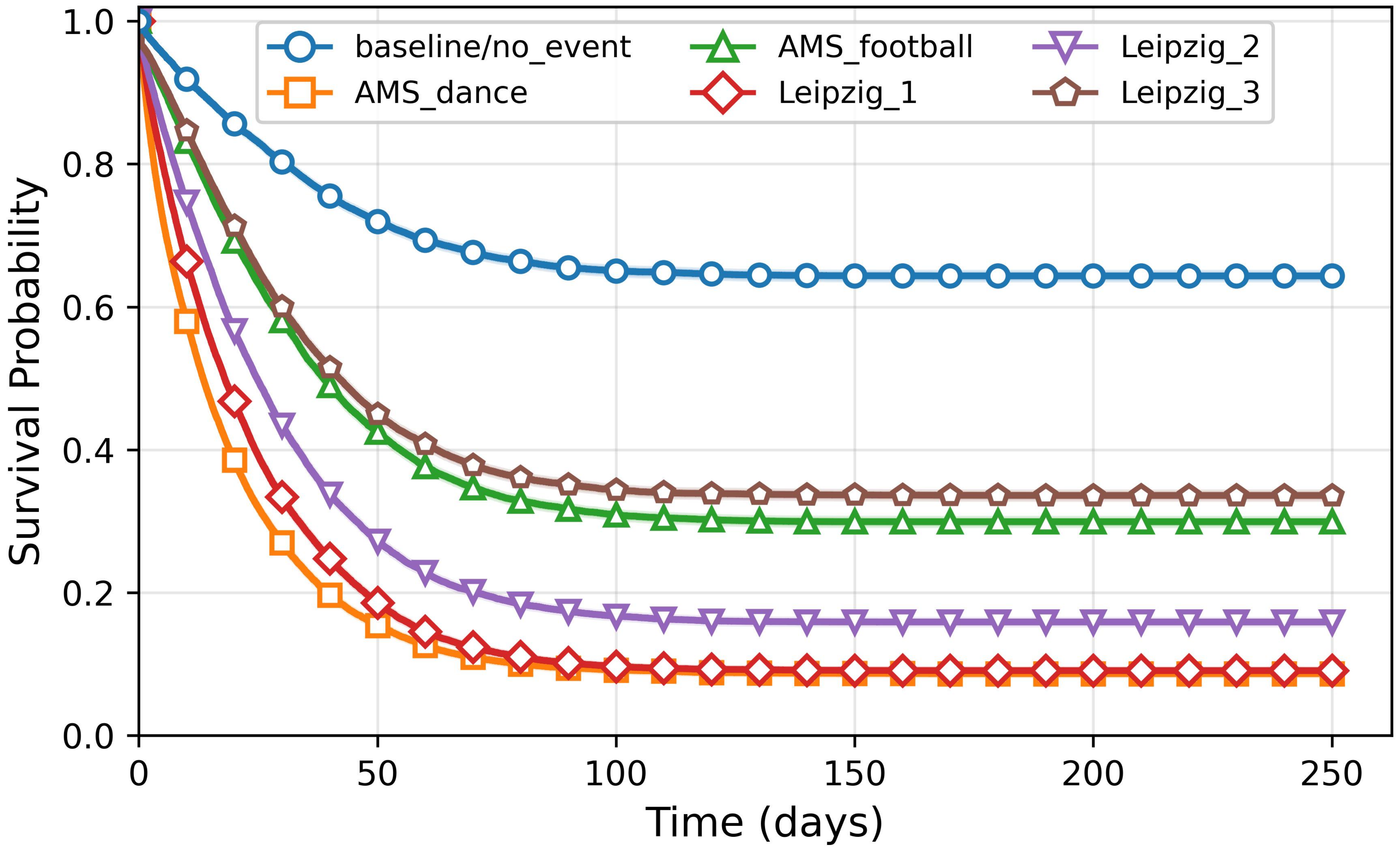}
\caption{Kaplan--Meier survival curves of district-level epidemic invasion under different mass gathering scenarios. Kaplan--Meier survival curves showing the probability that districts remain free of infectious individuals for the baseline (no-event) scenario and five representative mass gathering scenarios (AMS\_dance, AMS\_football, Leipzig\_1, Leipzig\_2, and Leipzig\_3). Earlier declines in survival probability indicate more rapid epidemic invasion.}
\label{fig:km-scenarios}
\vspace{-0.2cm}	
\end{figure}

Fig.~\ref{fig:km-scenarios} shows the Kaplan--Meier survival curves for districts remaining free of infectious individuals under the baseline and five MGE scenarios. The curves for all MGE scenarios decline earlier than the baseline curve, indicating earlier epidemic invasion in the presence of an MGE. The global log-rank test identified significant differences among the six survival distributions ($\chi^2=3787.8$, $df=5$, $p<0.001$), and each MGE scenario differed significantly from the baseline after correction for multiple comparisons. The magnitude of the difference varied across scenarios. For AMS\_dance, invasion was recorded in 19,180 of 21,000 district-simulation observations, compared with 7,490 under the baseline scenario. These results indicate that the association between MGE presence and epidemic invasion varies across event scenarios.

\begin{figure*}
\centering
\includegraphics[width=0.8\textwidth]
{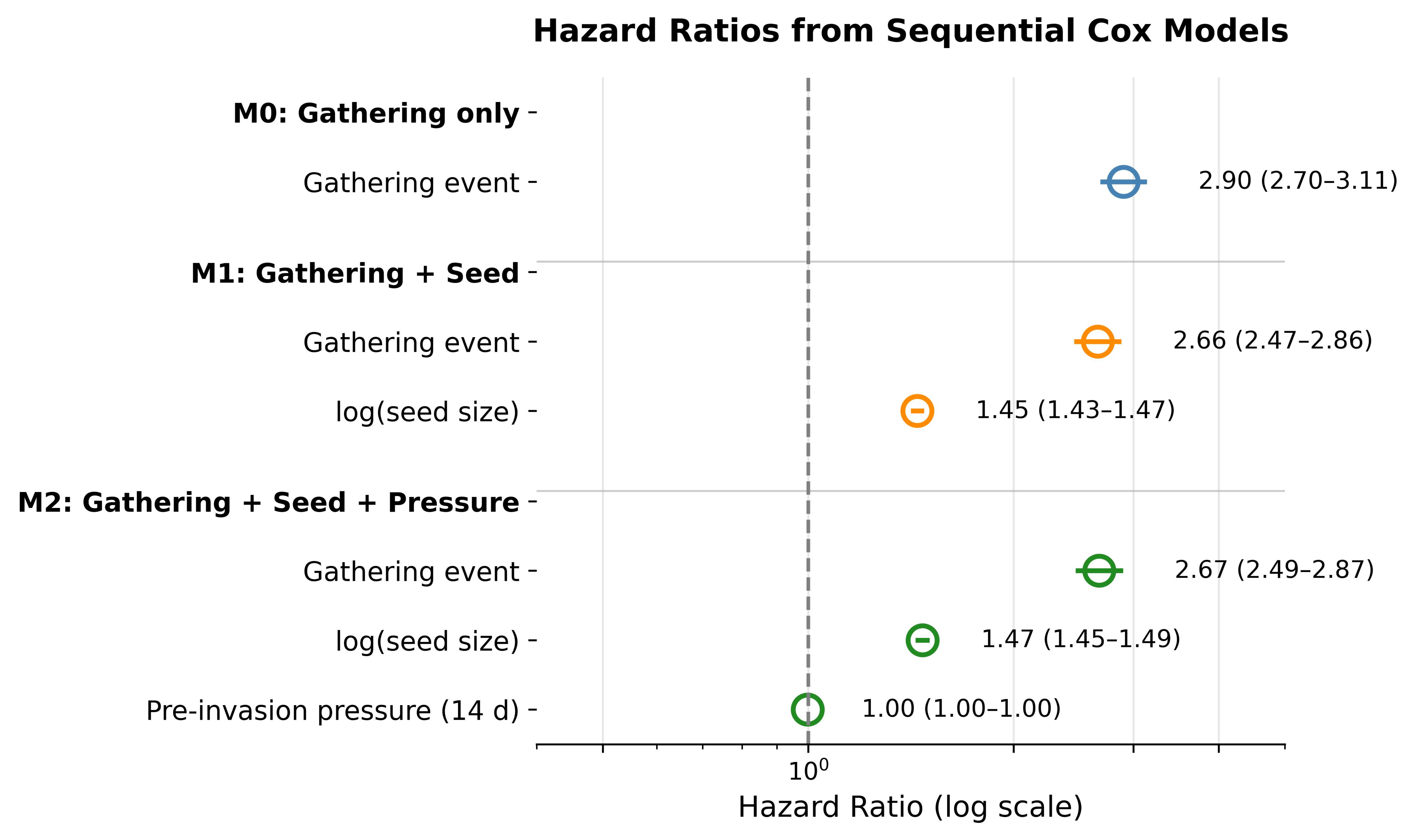}
\caption{Hazard ratios (HRs) and 95\% confidence intervals from sequential Cox proportional hazards models. M0 includes the MGE indicator; M1 additionally adjusts for log-transformed event-generated epidemic seed size; M2 further adjusts for mean pre-invasion transmission pressure. The dashed line indicates HR=1.}
\vspace{-0.2cm}	
\label{fig:forest-plot}
\end{figure*}

To quantify the association between MGEs and epidemic arrival timing, we fitted three sequential Cox proportional hazards models, see Fig.~\ref{fig:forest-plot}. In Model M0, MGE presence was associated with a higher hazard of district-level epidemic invasion (HR $=2.90$, 95\% CI $2.70$--$3.11$, $p<0.001$). After adjustment for the logarithm of the event-generated epidemic seed size in Model M1, the MGE hazard ratio decreased to 2.66 (95\% CI $2.47$--$2.86$, $p<0.001$). Larger event-generated epidemic seeds were also associated with earlier invasion (HR $=1.45$ per unit increase in log-transformed seed size, 95\% CI $1.43$--$1.47$, $p<0.001$). The attenuation of the MGE estimate suggests that differences in event-generated seed size statistically account for part of the observed association. However, the MGE association remained substantial after adjustment, indicating that differences in seed size alone did not fully account for the difference in invasion timing between MGE and baseline simulations.

Model M2 additionally incorporated mean daily transmission pressure over the 3-, 7-, or 14-day period preceding district invasion. The MGE and seed-size estimates remained broadly stable, at approximately HR $=2.67$ and HR $=1.47$, respectively. Model performance improved with the inclusion of pre-invasion transmission pressure, with the 14-day specification performing best among the windows considered. Its C-index was 0.634, compared with 0.575 for M0 and 0.618 for M1. Relative to the 14-day specification, $\Delta$AIC was 9,611 for M0, 1,435 for M1, 176 for the 3-day specification, and 75 for the 7-day specification. Full model comparisons are provided in the Supplementary Materials. Because transmission pressure was defined retrospectively relative to the observed arrival time, its contribution should be interpreted as an association with the pre-invasion transmission environment rather than evidence of prospective predictive performance or a causal effect.

Sensitivity analyses further examined the robustness of these associations. In the AMS\_dance scenarios, the association between event-specific transmission rate ($\beta_{\mathrm{event}}$) and invasion timing weakened substantially after adjustment for event-generated seed size, whereas the association with community transmissibility ($R_0$) remained pronounced. Across three initial seed districts, the adjusted MGE association remained positive, although its magnitude varied by seed location. Detailed parameter sensitivity and seed-location robustness results are reported in the Supplementary Materials.

% Conceptual interpretation
\begin{figure*}
\centering
%%%%%%%%%%%%%%%%%%%%%%%%%%%%%%%%%%%%%%%%%%%%%%%%%%%%%%%%%%%%%
%% Panel (a)
%%%%%%%%%%%%%%%%%%%%%%%%%%%%%%%%%%%%%%%%%%%%%%%%%%%%%%%%%%%%%
\begin{minipage}{12cm}
\centering
\vspace{0.0cm}
\begin{tikzpicture}[
>=Latex,
node distance=2.5cm,
box/.style={
draw,
rounded corners=4pt,
minimum width=2.8cm,
minimum height=1.0cm,
align=center,
fill=blue!12,
font=\bfseries\large
}
]

\node[box] (G) {Gathering};

\node[box, right=of G, fill=green!18] (I)
{Invasion\\Timing};

\draw[line width=1.4pt,->]
(G)--node[above,font=\bfseries]
{HR = 2.90}(I);
\end{tikzpicture}		
\end{minipage}
\vspace{0.2cm}

\text{(a) Model M0: Total association}

\hfill

%%%%%%%%%%%%%%%%%%%%%%%%%%%%%%%%%%%%%%%%%%%%%%%%%%%%%%%%%%%%%
%% Panel (b)
%%%%%%%%%%%%%%%%%%%%%%%%%%%%%%%%%%%%%%%%%%%%%%%%%%%%%%%%%%%%%
%\begin{minipage}{0.50\textwidth}
\begin{minipage}{14cm}
\centering		
\vspace{0.0cm}
\begin{tikzpicture}[
>=Latex,
node distance=2.5cm,
box/.style={
draw,
rounded corners=4pt,
minimum width=2.8cm,
minimum height=1.0cm,
align=center,
font=\bfseries\large
}
]

\node[box,fill=blue!12] (G)
{Gathering};

\node[box,fill=orange!25,right=of G] (S)
{Seed Size\\(log scale)};

\node[box,fill=green!18,right=of S] (I)
{Invasion\\Timing};

\draw[line width=1.5pt,blue,->]
(G)--(S);

\draw[line width=1.5pt,orange,->]
(S)--node[above,font=\bfseries]
{HR = 1.45}(I);

\draw[dashed,line width=1.3pt,gray!80,->]
(G) to[bend left=25]
node[above,font=\bfseries]
{HR = 2.66}
(I);

\end{tikzpicture}
\end{minipage}
\vspace{0.2cm}

\text{(b) Model M1: After adjustment for epidemic seed size}	

\vspace{0.0cm}	
\caption{Conceptual interpretation of the association between mass gatherings, event-generated epidemic seeding, and invasion timing. (a) Model M0 estimates the overall association between mass gathering events and district-level epidemic invasion. Gathering events are associated with substantially earlier epidemic invasion (HR = 2.90). (b) Model M1 additionally adjusts for the logarithm of the event-generated epidemic seed size. The gathering-event hazard ratio decreases to HR = 2.66, while larger epidemic seeds are independently associated with earlier invasion (HR = 1.45 per unit increase in log-transformed seed size). The reduction of the gathering-event estimate indicates that the seed size accounts for part of the observed association, but the remaining association indicates that seed size alone does not account for the difference in invasion timing between gathering and no-event simulations.}
\label{fig:conceptual-mechanism}
\end{figure*}

Fig.~\ref{fig:conceptual-mechanism} illustrates our conceptual interpretation of the association between MGE presence, epidemic seed size, and district-level epidemic invasion timing. Model M0 shows that MGE scenarios are associated with earlier epidemic invasion (HR = 2.90). After adjustment for the logarithm of the event-generated epidemic seed size in Model M1, the estimated MGE hazard ratio decreases to 2.66, while larger epidemic seeds are independently associated with earlier invasion (HR = 1.45). The reduction of the MGE hazard ratio after adjustment indicates that differences in the seed size account statistically for part of the observed association between MGE presence and invasion timing. However, the remaining association indicates that seed-size differences alone do not fully account for the observed difference in invasion timing. It is noted that other aspects of event-specific contact patterns and community transmission dynamics may also contribute to this association. These results should be interpreted as statistical associations within the simulated environment rather than evidence of direct causal pathways.

\section{Conclusion}
\label{sec:conclusion}
We developed a hybrid simulation framework coupling individual-based contact networks for mass gathering events (MGEs) with a stochastic metapopulation susceptible-exposed-infected-recovered (SEIR) model to investigate how localized, transient transmission events influence epidemic dynamics in urban environments. Using population and commuting data from Madrid, Spain, and contact networks representing different gathering scenarios, the framework links event-generated infections to subsequent epidemic progression and district-level invasion across a metropolitan mobility network.

The simulations yielded three main findings. First, MGEs were associated with larger epidemic peaks and faster epidemic progression, although the magnitude of these effects varied across event scenarios and levels of background transmissibility. Relative amplification was greatest at intermediate values of the basic reproduction number $R_0$, where event-generated infections could substantially alter subsequent epidemic trajectories. Scenarios with denser contact patterns exhibited stronger amplification, whereas measures reducing close-contact opportunities attenuated these effects.

Second, MGEs advanced district-level epidemic arrival and increased the temporal synchronization of invasion across the metropolitan area. The magnitude of these changes varied across event scenarios and districts, reflecting differences in event contact patterns and the spatial progression of transmission through the urban mobility network. The strongest advances and synchronization were observed in the AMS\_dance and AMS\_football scenarios, while stronger contact mitigation in the Leipzig scenarios produced arrival-time patterns closer to the baseline. These findings indicate that the metropolitan consequences of a localized gathering depend not only on transmission during the event but also on how the resulting infections propagate through the wider urban system.

Third, the sequential Cox models helped characterize the association between MGEs and earlier district-level epidemic invasion. Adjusting for the logarithm of event-generated epidemic seed size reduced the estimated MGE-associated hazard ratio from 2.90 to 2.66, but a substantial association remained. The association between the event-specific transmission rate $\beta_{\mathrm{event}}$ and invasion timing also weakened after adjustment for seed size, whereas the association with community transmissibility remained pronounced. These patterns are consistent with event-generated seed size accounting statistically for part, but not all, of the association between MGEs and earlier invasion. Incorporating mean daily transmission pressure over the preceding 14 days further improved model performance relative to the other specifications considered. Because this pressure measure was defined retrospectively relative to district invasion, its contribution reflects an association with the pre-invasion transmission environment rather than prospective predictive performance. More broadly, the Cox analyses characterize statistical associations within the simulated trajectories and do not independently identify causal pathways.

These findings suggest that limiting transmission during mass gatherings may have consequences beyond the event itself by reducing the number of infections introduced into the wider urban system and potentially slowing subsequent epidemic progression. The proposed framework provides a means of comparing event-management scenarios under different epidemiological conditions and examining how event-level contact patterns interact with community transmission and urban mobility. However, several limitations should be considered when interpreting the results. Infection risk during events is represented primarily through proximity-based contacts, without explicit modeling of environmental transmission factors such as ventilation and airflow. The model also assumes fixed mobility patterns and epidemiological parameters, does not explicitly represent behavioral adaptation or individual heterogeneity in susceptibility, immunity, and intervention compliance, and considers isolated rather than repeated or closely spaced events. Applying the framework to specific settings will therefore require careful consideration and validation of its assumptions. % Furthermore, the seed-location sensitivity analysis covered three representative initial epidemic seed districts rather than the full range of possible origins. 

Overall, this study demonstrates how transient, event-level contact patterns can shape metropolitan epidemic magnitude and spatial invasion through their interaction with event-generated epidemic seeding, community transmissibility, and urban mobility. By integrating event-scale contact networks, stochastic epidemic dynamics, and district-level survival analysis, the framework offers a multiscale approach to investigating the potential consequences of mass gatherings and comparing event-management strategies. Future work should extend and validate the framework across additional cities, pathogens, event types, and intervention scenarios to assess the generalizability of these findings and support context-specific epidemic risk assessment.

\section*{Acknowledgments}
This research is supported by the Singapore Ministry of Health’s National Medical Research Council under its National Epidemic Preparedness and Response R\&D Funding Initiative (MOH-001041) Programme for Research in Epidemic Preparedness And REsponse (PREPARE). This research is also partly supported by the Ministry of Education, Singapore, under its MOE AcRF Tier 2 Award MOET2EP20223-0003.

% bibliography
\bibliographystyle{IEEEtran}
\bibliography{refs_20261030a_sorted}

\vfill
\clearpage

\title{Supplementary Materials for:\\Hybrid epidemic simulation framework coupling equation-based and individual-based models}

\author{
Jaeyoung Kwak~\orcidlink{0000-0002-9627-3569}, 
Michael H. Lees~\orcidlink{0000-0002-5457-9180},
Chin Chun Ooi~\orcidlink{0000-0003-4813-4529}, and
Wentong Cai~\orcidlink{0000-0002-0183-3835}, \IEEEmembership{Senior Member, IEEE}
\thanks{Jaeyoung Kwak and Wentong Cai are with College of Computing and Data Science, Nanyang Technological University, Singapore.} % (e-mail: jaeyoung.kwak@ntu.edu.sg; ASWTCAI@ntu.edu.sg)
\thanks{Michael H. Lees is with Informatics Institute, University of Amsterdam, The Netherlands.} % (e-mail: M.H.Lees@uva.nl)
\thanks{Chin Chun Ooi is with Institute of High Performance Computing, Agency for Science, Technology and Research, Singapore.} % (e-mail: ooicc@a-star.edu.sg)
}

% \thanks{Manuscript received April 19, 2026; revised August 16, 2021.}}

% \markboth{Journal of \LaTeX\ Class Files,~Vol.~00, No.~0, Month~YEAR}% {Shell \MakeLowercase{\textit{et al.}}: A Sample Article Using IEEEtran.cls for IEEE Journals}

% Remember, if you use this you must call \IEEEpubidadjcol in the second column for its text to clear the IEEEpubid mark.

\maketitle

% \begin{abstract} % 100–250 word abstract
% \end{abstract}
% \begin{IEEEkeywords}
% hybrid epidemic simulation, metapopulation model, individual-based model, epidemic arrival time, mass gathering 
% \end{IEEEkeywords}

%=========================================================
% SUPPLEMENTARY MATERIALS
%=========================================================

% If compiled as a separate supplementary document, the following commands provide S-numbering for figures, tables, and equations.
\renewcommand{\thesection}{\Alph{section}}
\renewcommand{\thefigure}{S\arabic{figure}}
\renewcommand{\thetable}{S\arabic{table}}
\renewcommand{\theequation}{S\arabic{equation}}

\setcounter{figure}{0}
\setcounter{table}{0}
\setcounter{equation}{0}

\section{Dataset}
\label{sec:supplementary-dataset}
The hybrid simulation framework integrates two distinct layers: macroscopic demographic and mobility data for the metapopulation layer, and microscopic contact networks for the event-scale transmission layer.

\begin{figure}
\centering
\includegraphics[width=8cm]{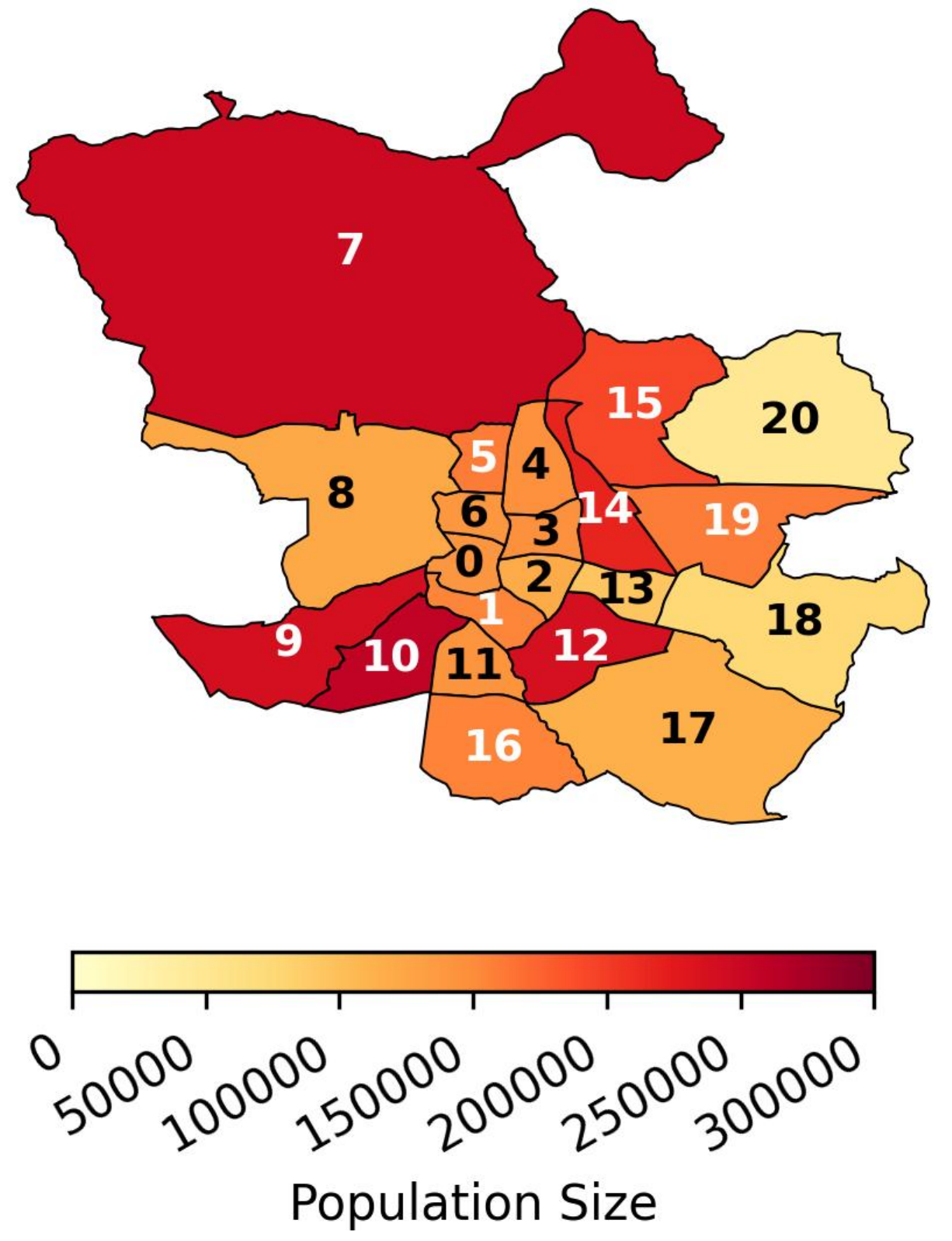}\vspace{+0.0cm}
\caption{Madrid district map.} 
\label{fig:madrid_district_map}
\end{figure}
\begin{table}
\normalsize                       %
\setlength{\tabcolsep}{6pt}       % general space between columns (6pt standard) 
\renewcommand{\arraystretch}{1.1} % general space between rows (1 standard)
\centering
\caption{Madrid population by district.}
\label{table:population_Madrid}
\resizebox{6.0cm}{!}{
\begin{tabular}{clr}
\hline
\textbf{District id} &
\textbf{District name} &
\textbf{Population} \\
\hline
\hline
0  & Centro             & 141,236 \\
1  & Arganzuela         & 154,243 \\
2  & Retiro             & 118,557 \\
3  & Salamanca          & 146,016 \\
4  & Chamartin          & 145,700 \\
5  & Tetuan             & 159,849 \\
6  & Chamberi           & 138,667 \\
7  & Fuencarral         & 247,692 \\
8  & Moncloa            & 120,834 \\
9  & Latina              & 240,155 \\
10 & Carabanchel        & 258,633 \\
11 & Usera              & 142,454 \\
12 & Puente De Vallecas & 239,057 \\
13 & Moratalaz          & 93,810 \\
14 & Ciudad Lineal      & 216,818 \\
15 & Hortaleza          & 193,228 \\
16 & Villaverde         & 154,808 \\
17 & Villa De Vallecas  & 114,733 \\
18 & Vicalvaro          & 75,485 \\
19 & San Blas           & 160,258 \\
20 & Barajas            & 50,077 \\
\hline
\hline
\end{tabular}%
}
\vspace{0.0cm}
\end{table}
\begin{figure}
\centering
\includegraphics[width=\columnwidth]{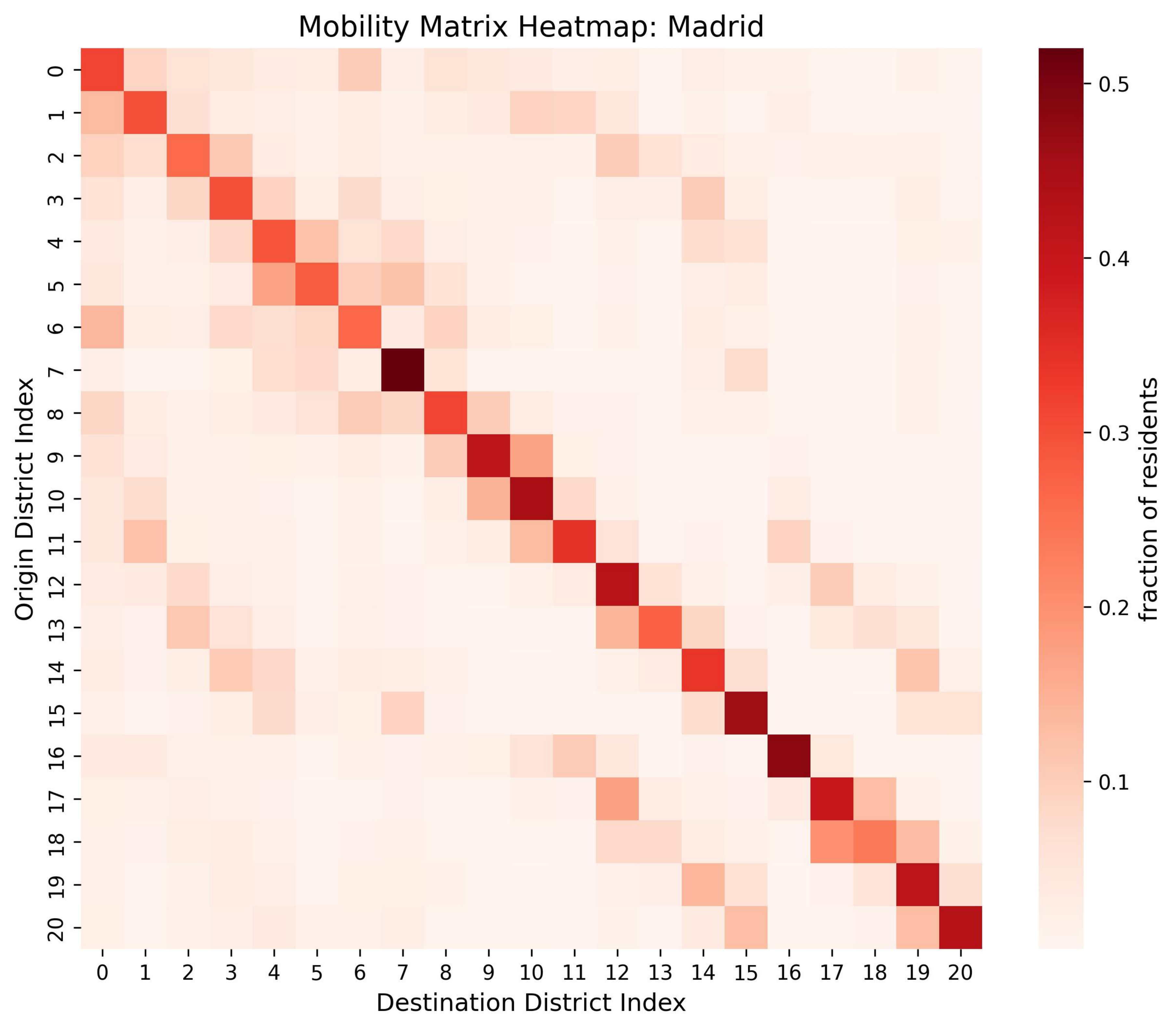}\vspace{-0.3cm}
\caption{Mobility matrix heatmap of inter-district commuting patterns in Madrid. Columns represent the fraction of residents from each district who commute to other districts, while rows indicate the corresponding inflow of commuters into each destination district. Darker red colors denote higher commuting volumes between district pairs. The pronounced diagonal elements indicate that the majority of residents remain within their home districts rather than commuting elsewhere on a daily basis.} 
\label{fig:mobility_matrix} 
\end{figure}

To simulate background community transmission, we utilize the urban structure of Madrid, Spain, based on population distributions and commuter flows described in Ref.~\cite{MiguelArribas_2023}. The study area comprises the 21 administrative districts of the Madrid municipality, see Fig.~\ref{fig:madrid_district_map}.
\begin{itemize}
\item {Population census:} A dataset defining the resident population $N_i$ for each district $i$. These values serve as the initial state for the subpopulation arrays. See Table~\ref{table:population_Madrid} for the population of Madrid at the district level.
\item {Origin-Destination (OD) matrix:} A matrix capturing daily commuting patterns. The raw flows are normalized to construct the mobility probability matrix $D_{ij}$, where each entry represents the probability that a resident of district $i$ commutes to district $j$ during the day-time step. See Fig.~\ref{fig:mobility_matrix} for the heat map of the mobility matrix of the commute patterns between districts in Madrid.
\end{itemize}

For individual-based transmission simulation, we use aggregated contact networks derived from empirical pedestrian trajectory data, which capture heterogeneous mixing patterns in real-world crowd environments. Rather than assuming homogeneous mixing, interactions are represented as a weighted contact network, where each edge corresponds to a unique pair of individuals $i$ and $j$, and the weight $\tau_{ij}$ denotes the cumulative duration of close contact within a 1.5\,m radius over the course of the event.

We evaluate the impact of mass gatherings using several empirical scenarios:
\begin{itemize}
\item {Amsterdam event data:} Data from dynamic crowd environments collected at the Johan Cruijff arena in Amsterdam, including a large indoor dance event ($N_{\text{event}} = 1048$) and a football match ($N_{\text{event}} = 362$)~\cite{Rutten_2022}. These datasets capture high-density interaction patterns under realistic event conditions. The dance event comprises three phases: a pre-event period (3.7\,h), a main event period (4\,h), and a post-event period (2\,h). During the main event, attendees predominantly remained stationary, whereas increased movement was observed before and after the main session, likely reflecting venue exploration and social interactions. The football match is divided into an entry phase (45\,min), first half (45\,min), halftime break (15\,min), second half (45\,min), and an extended exit phase (90\,min). Attendees were mostly seated during the two halves of the match, while substantially higher mobility occurred during entry, halftime, and exit periods.
\item {Leipzig experiment data:} Data from controlled concert experiments conducted in Leipzig, Germany, each involving approximately 1,100 participants~\cite{Moritz_2021}. The three experimental settings represent progressively stronger contact mitigation measures: (Leipzig\_1) a pre-pandemic baseline with unrestricted entry and exit through two main gates and fully occupied seating; (Leipzig\_2) a moderate mitigation scenario with quadrant-based access control, restricted movement between quadrants, and a checkerboard seating arrangement; and (Leipzig\_3) a high-mitigation scenario with paired seating separated by at least 1.5\,m and more number of entry and exit points are utilized. These scenarios provide a systematic basis for examining how crowd management and distancing measures alter contact patterns and transmission risk. All scenarios followed the same schedule, consisting of an entry period (60\,min), first half (20\,min), halftime break (20\,min), second half (20\,min), and exit phase (15\,min).
\end{itemize}

Table~\ref{table:ibm-data} provides the summary statistics for these datasets.

\begin{table*}
\normalsize                       %
\setlength{\tabcolsep}{8pt}       % general space between columns (6pt standard) 
\renewcommand{\arraystretch}{1.2} % general space between rows (1 standard)
% \centering
\caption{Summary statistics for the individual-based model (IBM) datasets. $N_{\text{event}}$ represents the total number of attendees. The mean nodal degree ($k_{\text{avg}}$) and its standard deviation ($k_{\text{std}}$) describe the distribution of unique contacts per person within the defined proximity threshold of 1.5\,m.}
\label{table:ibm-data}
%\resizebox{12.0cm}{!}{
%\resizebox{1.1\textwidth}{!}{
\centering
\resizebox{12.0cm}{!}{
\begin{tabular}{llrrrrc}
	\hline\hline
	Scenario & Location & $N_{\text{event}}$ & $k_{\text{avg}}$ & $k_{\text{std}}$ & Refs. \\
	\hline
	Football match (AMS\_football)& Amsterdam, NL & 362 & 17.3 & 9.4 & \cite{Rutten_2022} \\
	Dance event (AMS\_dance)& Amsterdam, NL & 1048 & 161.7 & 50.0 & \cite{Rutten_2022} \\
	Indoor concert scenario 1 (Leipzig\_1) & Leipzig, DE & 1194 & 35.6 & 11.2 & \cite{Moritz_2021} \\
	Indoor concertscenario 2 (Leipzig\_2) & Leipzig, DE & 1158 & 18.5 & 7.2 & \cite{Moritz_2021} \\
	Indoor concertscenario 3 (Leipzig\_3) & Leipzig, DE & 1054 & 8.7 & 4.2 & \cite{Moritz_2021} \\
	\hline\hline
\end{tabular}
}
\vspace{-0.5cm}
\end{table*}

\section{Stochastic metapopulation SEIR model}
\label{sec:supplementary-SEIR}
The foundation of our simulation framework is a stochastic metapopulation model that describes the spread of an epidemic across a network of $n$ geographically distinct patches, or districts. The model captures community-level transmission while representing daily mobility between districts.

The total population is disaggregated into home--destination subpopulations $N_{ij}$, where the first index $i$ denotes an individual's home district and the second index $j$ denotes the district in which the individual is located during the away-from-home phase. The simulation proceeds in discrete time steps of $\Delta t=0.5$~days. Each calendar day therefore consists of two consecutive phases: an \emph{away-from-home phase} (hereafter, the \emph{away phase}), during which individuals travel to and interact in their destination districts, and a \emph{home phase}, during which individuals return to their home districts.

Mobility is governed by a static matrix $D_{ij}$, where $D_{ij}$ denotes the probability that an individual residing in district $i$ travels to district $j$ during the away phase. During the away phase, each district therefore contains a heterogeneous mixture of residents and commuters from different home districts. During the subsequent home phase, individuals return to their home districts, such that transmission occurs within the corresponding home-district populations.

Within each subpopulation $N_{ij}$, individuals occupy one of four epidemiological compartments: Susceptible ($S_{ij}$), Exposed ($E_{ij}$), Infectious ($I_{ij}$), and Recovered ($R_{ij}$).

The force of infection is defined locally for each district and simulation phase. Let $p\in{A,H}$ denote the away and home phases, respectively. For a generic district $\ell$, the phase-specific force of infection is
\begin{equation}
\lambda_{\ell}^{p}(t)
=
\beta\frac{I_{\ell}^{p}(t)}{N_{\ell}^{p}(t)},
\qquad
p\in{A,H},
\label{eq:infection-force-general}
\end{equation}
\noindent
where $\beta$ is the community transmission rate, and $I_{\ell}^{p}(t)$ and $N_{\ell}^{p}(t)$ denote the numbers of infectious individuals and total individuals present in district $\ell$ during phase $p$, respectively. The superscript identifies the phase in which the force of infection is evaluated, whereas the subscript identifies the district.

During the away phase, individuals are grouped according to their current destination district and their size is given as
\begin{align}
N_j^{A}(t) &= \sum_{i=1}^{n} N_{ij}(t),\\
I_j^{A}(t) &= \sum_{i=1}^{n} I_{ij}(t),
\end{align}
\noindent
and the corresponding force of infection in destination district $j$ is
\begin{equation}
\lambda_j^{A}(t) = \beta\frac{I_j^{A}(t)}{N_j^{A}(t)}.
\label{eq:infection-force-away}
\end{equation}

During the home phase, individuals are grouped according to their home district and their size is given as
\begin{align}
N_i^{H}(t) &= \sum_{j=1}^{n} N_{ij}(t),\\
I_i^{H}(t) &= \sum_{j=1}^{n} I_{ij}(t),
\end{align}
\noindent
and the corresponding force of infection in home district $i$ is
\begin{equation}
\lambda_i^{H}(t) = \beta\frac{I_i^{H}(t)}{N_i^{H}(t)}.
\label{eq:infection-force-home}
\end{equation}

Thus, $\lambda_j^{A}(t)$ and $\lambda_i^{H}(t)$ represent the force of infection during the two types of half-day simulation phases. The same home--destination subpopulations are aggregated differently in the two phases: by current destination district during the away phase and by home district during the home phase.

Transitions between epidemiological compartments are modeled as stochastic events. During each simulation phase of duration $\Delta t$, the number of individuals transitioning between compartments is drawn from a binomial distribution~\cite{Tizzoni_2014,Wardle_2023}. For susceptible individuals in home--destination subpopulation $(i,j)$, the probability of becoming exposed during a given phase is
\begin{equation}
p_{ij}^{E}(t) = 1-\exp\left[-\lambda_{ij}(t)\Delta t\right],
\label{eq:exposure-probability}
\end{equation}
\noindent
where the phase-specific force of infection experienced by the subpopulation is
\begin{equation}
\lambda_{ij}(t)
=
\begin{cases}
\lambda_j^{A}(t), & \text{during the away phase},\\
\lambda_i^{H}(t), & \text{during the home phase}.
\end{cases}
\label{eq:subpopulation-foi}
\end{equation}
\noindent
The number of new exposures is then
\begin{equation}
\Delta E_{ij}(t) \sim \mathrm{Binomial} \left( S_{ij}(t), p_{ij}^{E}(t) \right).
\label{eq:new-exposures}
\end{equation}

Transitions from the exposed to infectious state and from the infectious to recovered state are modeled analogously:
\begin{align}
\Delta I_{ij}(t)
&\sim
\mathrm{Binomial}
\left( E_{ij}(t), 1-\exp(-\alpha\Delta t) \right),
\\
\Delta R_{ij}(t)
&\sim
\mathrm{Binomial}
\left( I_{ij}(t), 1-\exp(-\gamma\Delta t) \right),
\end{align}
\noindent
where
\begin{equation}
\alpha=T_{\mathrm{inc}}^{-1},
\qquad
\gamma=T_{\mathrm{inf}}^{-1},
\end{equation}
\noindent
are the progression and recovery rates, defined as the reciprocals of the mean incubation period $T_{\mathrm{inc}}$ and infectious period $T_{\mathrm{inf}}$, respectively.

For analyses requiring a transmission-pressure measure at a daily rather than half-day temporal resolution, the two phase-specific forces of infection are aggregated into a daily force-of-infection measure. Let calendar day $k$ consist of an away phase followed by a home phase. Because the two phases have equal duration, the daily transmission pressure in district $i$ is defined as
\begin{equation}
\Lambda_i(k)
=
\frac{1}{2}
\left[
\lambda_i^{A}(k)
+
\lambda_i^{H}(k)
\right],
\label{eq:daily-infection-force}
\end{equation}
\noindent
where $\lambda_i^{A}(k)$ and $\lambda_i^{H}(k)$ denote the phase-specific forces of infection during the away and home phases of calendar day $d$, respectively. The daily quantity $\Lambda_i(k)$ therefore represents the average transmission pressure experienced in district $i$ across the two half-day phases comprising one calendar day. It is distinct from the phase-specific force of infection $\lambda_{\ell}^{p}(t)$, which is directly used to determine stochastic SEIR transitions within each simulation phase.

Table~\ref{table:notation_SEIR} shows a list of model parameters and notation used in the stochastic metapopulation SEIR model.

\begin{table*}
\normalsize                       %
\setlength{\tabcolsep}{8pt}       % general space between columns (6pt standard) 
\renewcommand{\arraystretch}{1.2} % general space between rows (1 standard)
% \centering
\caption{Parameters and notation used in the stochastic metapopulation SEIR model}
\label{table:notation_SEIR}
\centering
\resizebox{12.0cm}{!}{
\begin{tabular}{ll}
	\hline\hline
	Symbol & Description \\
	\hline
	\(n\) & Number of districts \\
	\(i\) & Home-district index \\
	\(j\) & Destination-district index \\
	\(\ell\) & Generic district index \\
	\(N_{ij}\) & Population whose home district is \(i\) and current destination is \(j\) \\
	\(D_{ij}\) & Probability of travel from district \(i\) to district \(j\) during the away phase \\
	\(S_{ij},E_{ij},I_{ij},R_{ij}\) & SEIR populations in home--destination subpopulation \((i,j)\) \\
	\(A,H\) & Away-from-home and home phases \\
	\(\Delta t\) & Duration of one simulation phase (\(0.5\) day) \\
	\(\beta\) & Community transmission rate \\
	\(\lambda_\ell^p\) & Phase-specific force of infection in district \(\ell\) during phase \(p\) \\
	\(\Lambda_i(k)\) & Daily average transmission pressure in district \(i\) \\
	\(\alpha\) & Exposed-to-infectious progression rate \\
	\(\gamma\) & Infectious-to-recovered rate \\
	\(T_{\mathrm{inc}}\) & Mean incubation period \\
	\(T_{\mathrm{inf}}\) & Mean infectious period\\
	\hline\hline
\end{tabular}
}
\vspace{-0.2cm}
\end{table*}

\section{Epidemic progression and final epidemic size}
\label{sec:supplementary-epidemic-progression}

\begin{figure}
\begin{minipage}[t]{\columnwidth}
(a)\\
\vspace{0.1cm}
\includegraphics[width=\linewidth]{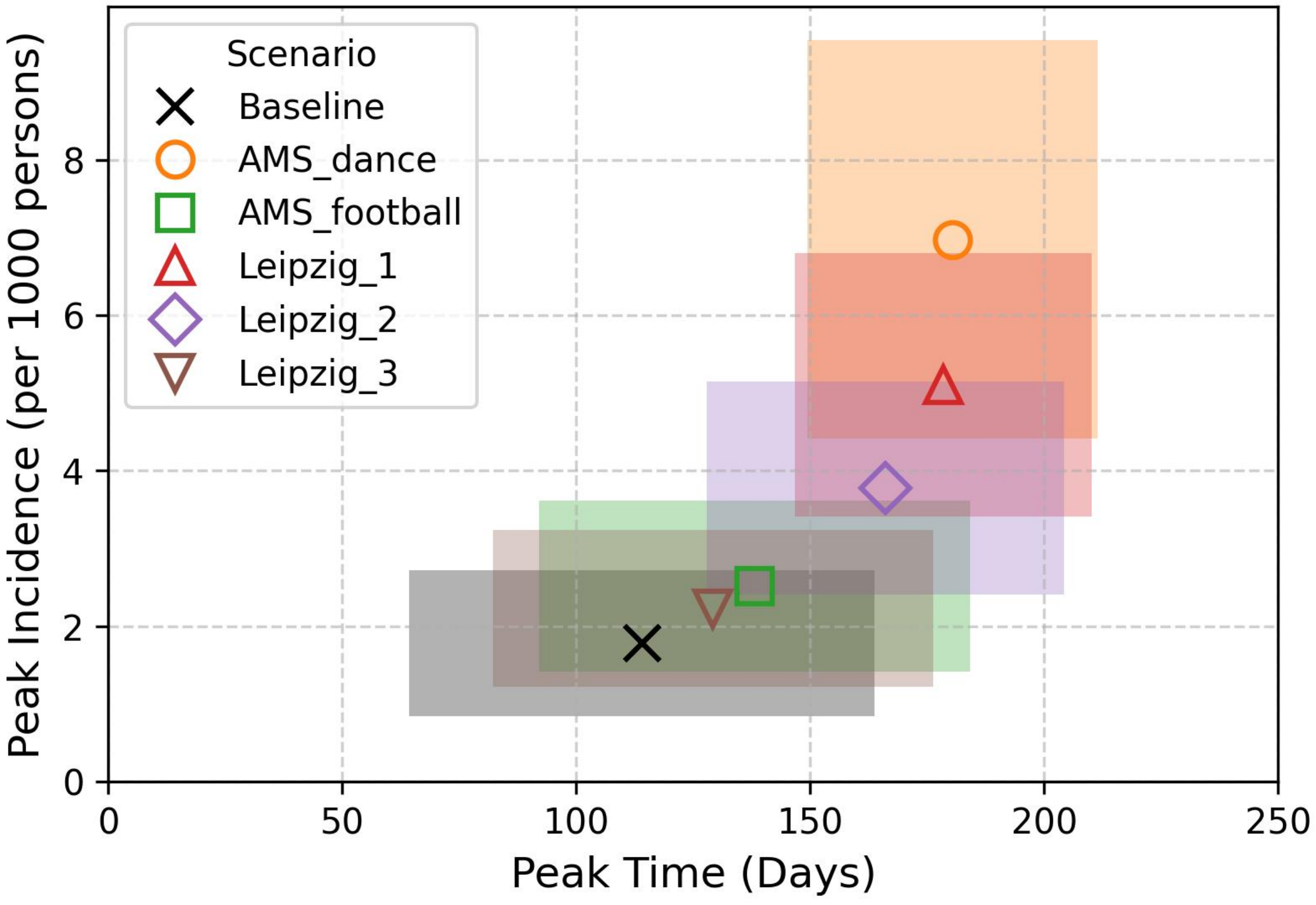}
\end{minipage}
\\
\begin{minipage}[t]{\columnwidth}
(b)\\
\vspace{0.1cm}	
\includegraphics[width=\linewidth]{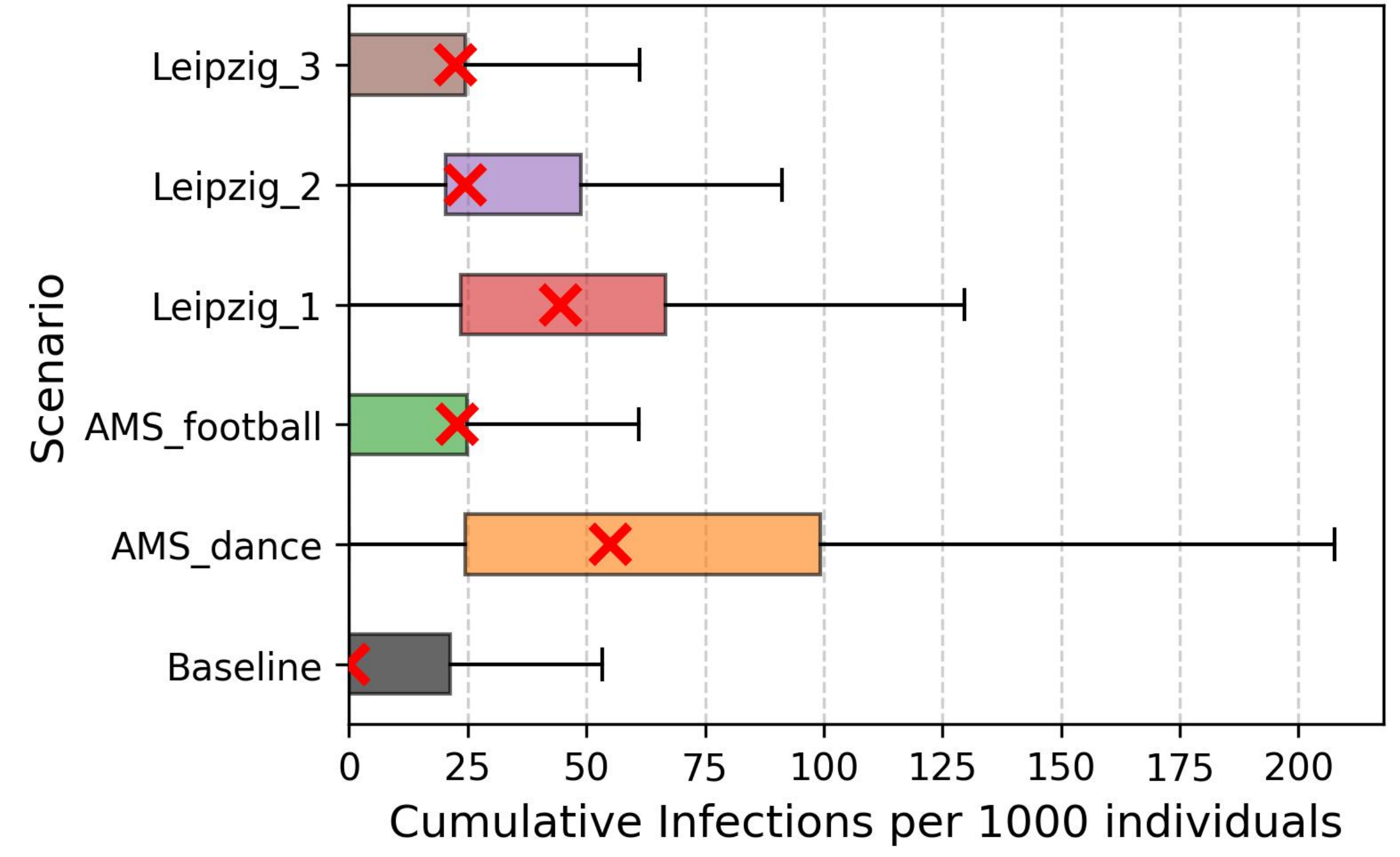}
\end{minipage}
\caption{Epidemic progression under different mass gathering event (MGE) scenarios: (a) Summary of active case curves: peak time and peak incidence (active cases per 1000 persons). Different symbols indicate the mean values of peak time and peak incidence for different MGE scenarios. Rectangle width and height indicate the standard deviation of peak time and peak incidence, respectively. (b) Boxplots visualizing the distribution of final epidemic size.}
\label{fig:supplementary-temporal-progression}
\vspace{-0.2cm}
\end{figure}

Figure~\ref{fig:supplementary-temporal-progression} summarizes the active-case trajectories across the baseline and mass gathering event (MGE) scenarios. Fig.~\ref{fig:supplementary-temporal-progression}(a) shows that higher epidemic peaks were generally associated with later peak times across the baseline and mass gathering event (MGE) scenarios. That is, scenarios with greater epidemic amplification did not necessarily reach their maximum earlier, despite exhibiting faster initial growth. Fig.~\ref{fig:supplementary-temporal-progression}(b) shows a similar association between epidemic seed size and final epidemic size, with scenarios generating larger seeds generally producing larger cumulative outbreaks. 

\begin{figure*}[t]
\centering
\includegraphics[width=14cm]{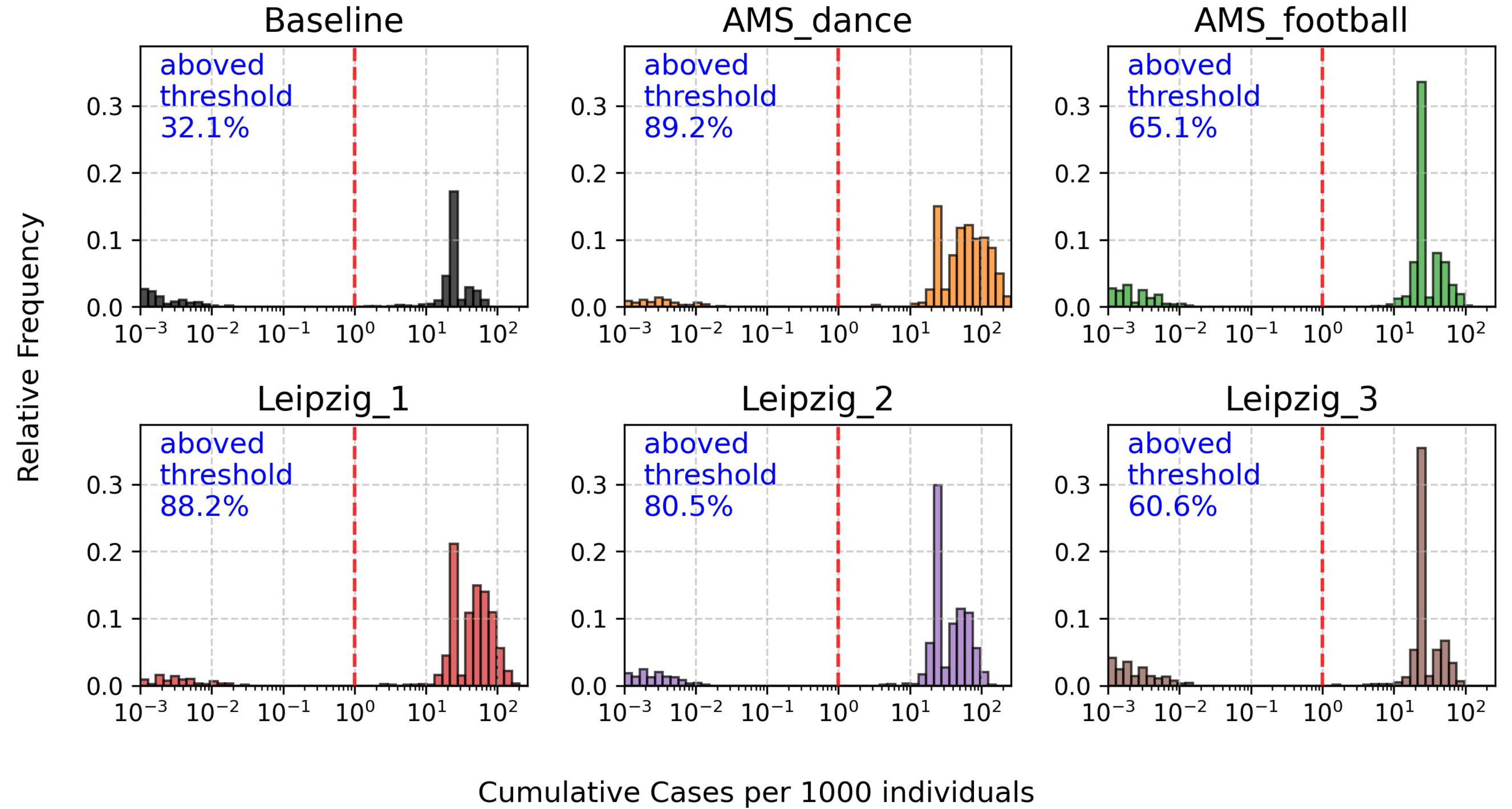}
\caption{Distribution of cumulative cases per 1,000 residents at day 250. Red dashed vertical lines indicate the threshold of one cumulative case per 1,000 residents. Results are based on 1,000 independent simulations with $R_0=1.5$, $\beta_{\mathrm{event}}=0.5$, and $I_{\mathrm{ss}}=1$.}
\label{fig:final-epidemic-size-distribution}
\end{figure*}

Figure~\ref{fig:final-epidemic-size-distribution} further characterizes the distribution of cumulative cases per 1,000 residents at day 250 across 1,000 independent stochastic simulations. The red dashed vertical lines indicate the threshold of one cumulative case per 1,000 residents. For the representative parameter setting ($R_0=1.5$, $\beta_{\mathrm{event}}=0.5$, and $I_{\mathrm{ss}}=1$), approximately 60--90\% of simulation runs under the MGE scenarios exceeded this threshold by day 250, compared with approximately 32\% under the baseline scenario. These results indicate that MGE scenarios were associated with a higher frequency of threshold exceedance. Within the Leipzig scenarios, stronger contact-mitigation measures were associated with less epidemic amplification and a lower frequency of threshold exceedance. Together, these results illustrate variation in both epidemic progression and cumulative epidemic burden across stochastic simulation runs.

\section{Seed-destination dependent epidemic arrival times}
\label{sec:supplementary-seed-destination-arrival}
This section characterizes the spatial dependence of epidemic arrival times on the initial epidemic seed location. Each of the 21 districts in Madrid was considered in turn as the seed district, and 1,000 independent stochastic simulation runs were performed for each seed location. For each run, the arrival time was recorded for every destination district.

Let \(i\) and \(j\) denote the epidemic seed and destination districts, respectively. The mean arrival time at destination \(j\), given seed district \(i\), is defined as
\begin{equation}
T_{ij}
=
\frac{1}{N_{\mathrm{runs}}}
\sum_{r=1}^{N_{\mathrm{runs}}}
T_{ij,r},
\label{eq:appendix_mean_arrival_time}
\end{equation}
\noindent
where \(T_{ij,r}\) is the arrival time at destination district \(j\) in simulation run \(r\), and \(N_{\mathrm{runs}}=1000\). The resulting \(21\times21\) matrix summarizes how epidemic arrival times vary with the locations of both the initial seed and the destination. Diagonal entries (\(i=j\)) are omitted from the visualizations because the seed district represents the initial condition rather than a subsequent epidemic arrival.

\begin{figure*}%[p]
\centering
\includegraphics[width=17cm]{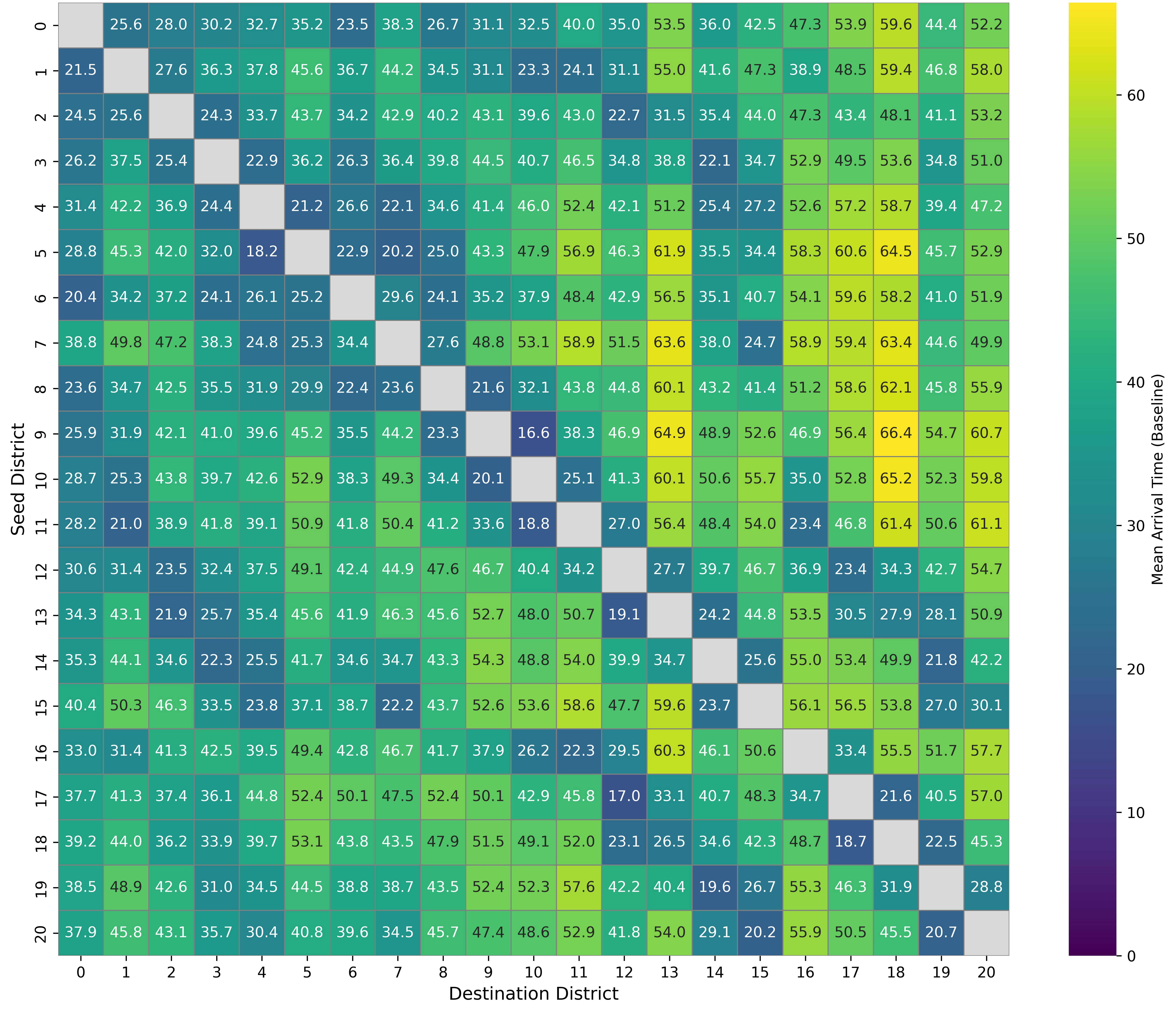}
\vspace{+0.3cm}
\caption{Mean epidemic arrival time for the baseline scenario. Rows indicate the initial epidemic seed district and columns indicate the destination district. Each cell represents the mean arrival time over 1,000 independent simulation runs. Diagonal entries are omitted because the destination coincides with the initial seed. Results correspond to \(R_0=1.5\) and \(\beta_{\mathrm{event}}=0.5\).}
\label{fig:arrival_matrix_baseline}
\end{figure*}

Fig.~\ref{fig:arrival_matrix_baseline} shows the mean arrival-time matrix for the baseline scenario, with seed districts in rows and destination districts in columns. The matrix exhibits substantial heterogeneity across seed--destination pairs, indicating that arrival time depends on the initial seed location as well as the destination. In particular, the same destination district can be reached at different times depending on where the epidemic originates.

\begin{figure*}%[p]
\centering
\includegraphics[width=17cm]{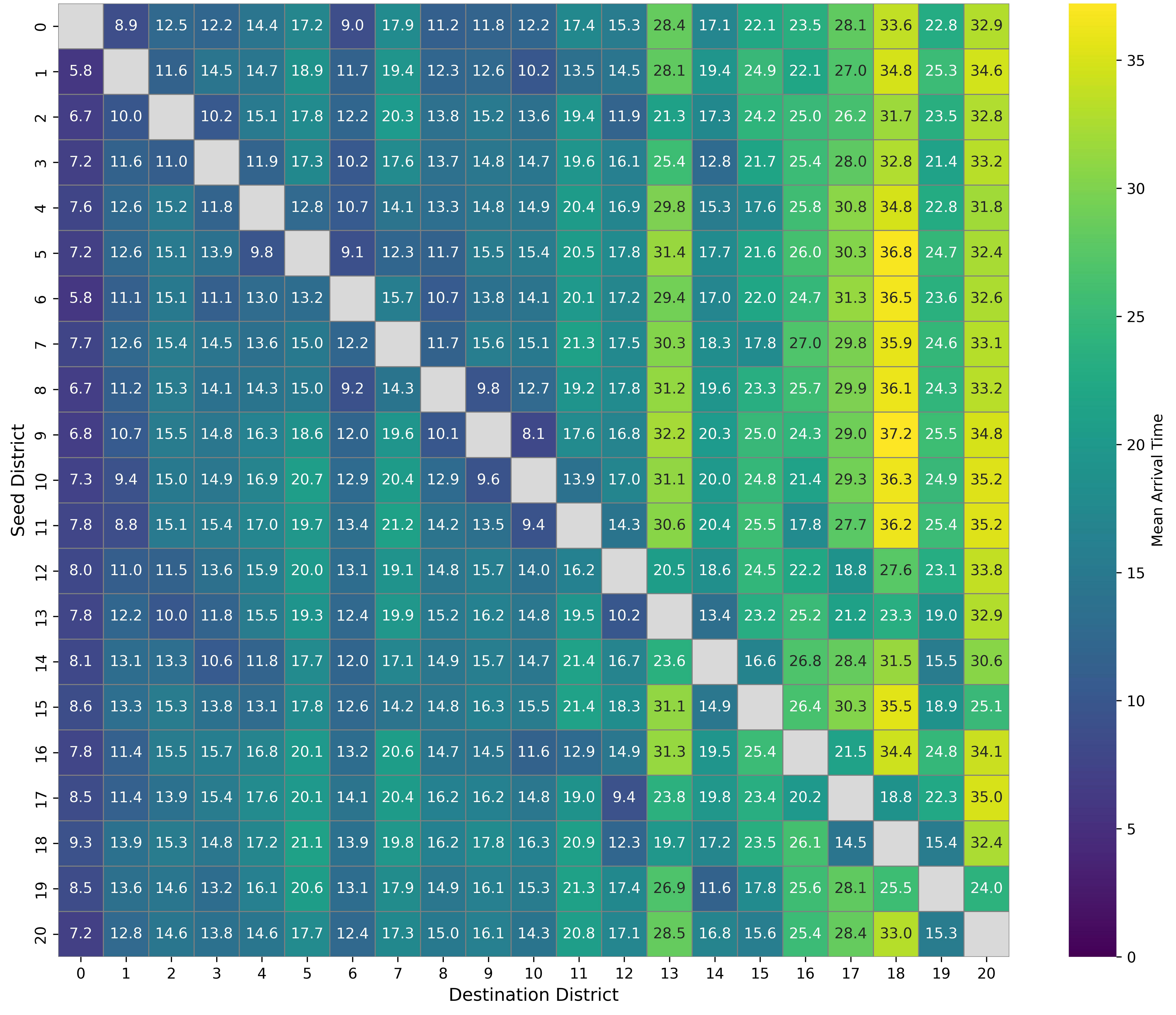}
\vspace{+0.3cm}
\caption{Mean epidemic arrival time for the AMS\_dance scenario. Rows indicate the initial epidemic seed district and columns indicate the destination district. Each cell represents the mean arrival time over 1,000 independent simulation runs. Diagonal entries are omitted because the destination coincides with the initial seed. Results correspond to \(R_0=1.5\) and \(\beta_{\mathrm{event}}=0.5\).}
\label{fig:arrival_matrix_ams_dance}
\end{figure*}

Figure~\ref{fig:arrival_matrix_ams_dance} presents the corresponding matrix for the AMS\_dance mass-gathering event (MGE) scenario, using the same values of \(R_0=1.5\) and \(\beta_{\mathrm{event}}=0.5\). Compared with the baseline scenario, the MGE changes the spatial pattern of epidemic arrival times, with the magnitude of the change varying across seed--destination pairs. Some pairs exhibit relatively small shifts, whereas others experience substantially earlier arrival. These differences suggest that the spatial impact of the event depends on the initial epidemic location and the destination district.

To quantify these changes, Fig.~\ref{fig:arrival_matrix_diff} shows the difference between the mean arrival times in the MGE and baseline scenarios for each seed--destination pair:
\[
\Delta T_{ij}
=
T_{ij,\mathrm{MGE}} - T_{ij,\mathrm{baseline}}.
\]
\noindent
Negative values indicate earlier mean arrival in the MGE scenario. The shifts vary substantially across the matrix, with some seed--destination pairs exhibiting advances exceeding 30 days. Thus, the effect of the MGE on epidemic arrival is spatially heterogeneous rather than uniform across the metropolitan area.

\begin{figure*}%[p]
\centering
\includegraphics[width=17cm]{figures/epidemic_arrival_time_matrix_diff.pdf}
\caption{Difference in mean epidemic arrival time between the AMS\_dance and baseline scenarios for each initial seed--destination pair. Values are defined as \(\Delta T_{ij}=T_{ij,\mathrm{MGE}}-T_{ij,\mathrm{baseline}}\), in days; negative values indicate earlier mean arrival in the AMS\_dance scenario. Results are averaged over 1,000 independent stochastic simulation runs for each pair, with \(R_0=1.5\), \(\beta_{\mathrm{event}}=0.5\), and \(I_{ss}=1\). Diagonal entries are omitted because the destination coincides with the initial seed.}
\label{fig:arrival_matrix_diff}
\end{figure*}

These matrices provide a seed--destination-level view of the spatial patterns underlying the city-wide arrival-time shifts. Although the AMS\_dance scenario generally advances epidemic arrival, the extent of this advancement depends on both the initial epidemic seed location and the destination district. The results therefore highlight the role of spatial configuration in shaping epidemic invasion across the metropolitan area.

\section{Cox proportional hazards models}
\label{sec:supplementary-cox-models}
District-level epidemic invasion timing was analyzed using Cox proportional hazards models. For district $i$ in simulation replicate $r$, the arrival time, $T_{ir}^{\mathrm{arr}}$, was defined as the first simulation time at which the district contained at least one infectious individual ($I_{ir}\geq 1$). Districts without an infectious individual by the end of the simulation horizon were treated as right-censored. Thus, each replicate contributed one survival record per district.

Districts within the same replicate share an epidemic realization, mobility patterns, and transmission history. Moreover, the MGE indicator and event-generated epidemic seed are shared by all districts in a replicate. To account for within-replicate dependence, we estimated cluster-robust (Huber--White sandwich) standard errors, using the simulation replicate as the clustering unit~\cite{Lin_1989}. Hazard ratios (HRs) greater than one indicate a higher instantaneous hazard of first invasion and therefore an association with earlier invasion, whereas HRs below one indicate an association with later invasion.

We modeled district-level epidemic invasion hazards using the Cox proportional hazards model~\cite{Lin_1989,Kleinbaum_2011},
\begin{equation}
h_i(t) = h_0(t) \exp\left( \mathbf{x}_i^{\top}\boldsymbol{\theta} \right),
\label{eq:cox-general}
\end{equation}
\noindent
where $h_i(t)$ denotes the hazard of epidemic invasion for district $i$ at time $t$, $h_0(t)$ is the unspecified baseline hazard, $\mathbf{x}_i$ is the vector of explanatory variables, and $\boldsymbol{\theta}$ is the corresponding vector of regression coefficients. Hazard ratios greater than one indicate an association with earlier epidemic invasion, whereas hazard ratios below one indicate an association with delayed invasion.

We fitted three nested Cox proportional hazards models:
\begin{align}
\text{M0:} \quad h_i(t)
&=
h_0(t) \exp\left( \theta_G G_i \right),
\\
\text{M1:} \quad h_i(t)
&=
h_0(t) \exp\left( \theta_G G_i + \theta_Z\log(Z_i+1) \right),
\\
\text{M2:} \quad h_i(t)
&=
h_0(t) \exp\left( \theta_G G_i + \theta_Z\log(Z_i+1) + \theta_{\Lambda}\bar{\Lambda}_i^{(W)} \right).
\end{align}
\noindent
Here, $G_i$ is a binary indicator denoting whether the stochastic simulation replicate includes a mass gathering event, and $Z_i$ denotes the event-generated epidemic seed, defined as the total number of exposed and infectious individuals generated directly by the gathering. The quantity $\bar{\Lambda}_i^{(W)}$ denotes the mean daily transmission pressure in district $i$ during the $W$ days preceding epidemic arrival. The logarithmic transformation of $Z_i$ reduces the influence of extremely large epidemic seeds while preserving their relative ordering.

M0 estimates the unadjusted association between MGE occurrence and invasion timing. M1 additionally adjusts for event-generated seed size, while M2 adds recent pre-invasion transmission pressure. The change in the MGE coefficient across these specifications was used to examine how the estimated association varied with adjustment for event-generated seeding and the pre-invasion transmission environment. These sequential models were intended primarily to characterize associations within the simulated trajectories rather than to establish causal pathways.

The daily transmission-pressure measure $\Lambda_i(k)$ is defined in the stochastic metapopulation SEIR model as the average of the away-from-home and home phase-specific forces of infection over simulation day $k$ (Eq.~\ref{eq:daily-infection-force}). To characterize the transmission environment immediately preceding epidemic invasion, we calculated the mean daily transmission pressure over a window of $W$ days before arrival:
\begin{equation}
\bar{\Lambda}^{(W)}_i 
= \frac{1}{W} \sum_{k=1}^{W} \Lambda_i \left(T^{\mathrm{arr}}_{i} - k\right)
\label{eq:avg-infection-force}
\end{equation}
\noindent
where $T_i^{\mathrm{arr}}$ denotes the epidemic arrival time in district $i$ expressed in calendar days, and $\Lambda_i(T_i^{\mathrm{arr}}-k)$ is the daily transmission pressure in district $i$ on the $k$~th day preceding epidemic arrival. Thus, $\bar{\Lambda}_i^{(W)}$ represents a temporally aligned summary of the recent transmission environment immediately before invasion. We evaluated averaging windows of $W=3$, $7$, and $14$ days to examine the sensitivity of the estimated association to the temporal scale over which recent transmission pressure was aggregated. These windows represent progressively broader transmission histories relative to the mean latent period of four days and infectious period of five days used in the simulation model.

Because $\bar{\Lambda}_i^{(W)}$ is defined retrospectively relative to the observed epidemic arrival time, it was used as an exploratory, temporally aligned measure of the transmission environment preceding invasion rather than as a prospectively available baseline predictor. Consequently, Model M2 was interpreted as an extension of the mechanistic analysis that assesses whether recent transmission pressure contains additional statistical information about invasion timing. It was not interpreted as evidence of a causal effect of pre-invasion transmission pressure.

\section{Performance of sequential Cox proportional hazards models}
\label{sec:supplementary-Cox-analysis-performance}
We compared the sequential Cox proportional hazards models using the Akaike Information Criterion (AIC) and Harrell's concordance index (C-index). The AIC was calculated as
\begin{equation}
\mathrm{AIC}=2q-2\log(L_{\mathrm{partial}}),
\end{equation}
\noindent
where $q$ is the number of estimated regression coefficients and $L_{\mathrm{partial}}$ is the maximized Cox partial likelihood. We report $\Delta$AIC relative to the model with the lowest AIC, with smaller values indicating a more favorable trade-off between model fit and complexity. The C-index measures the agreement between predicted risk rankings and observed event-time orderings among comparable pairs, accounting for right-censored observations. Values of 0.5 and 1.0 indicate chance-level and perfect discrimination, respectively. The C-index was used to assess within-sample discrimination, not out-of-sample predictive performance.

\begin{figure}
\begin{minipage}[t]{\columnwidth}
(a)\\
\vspace{0.1cm}
\includegraphics[width=\linewidth]{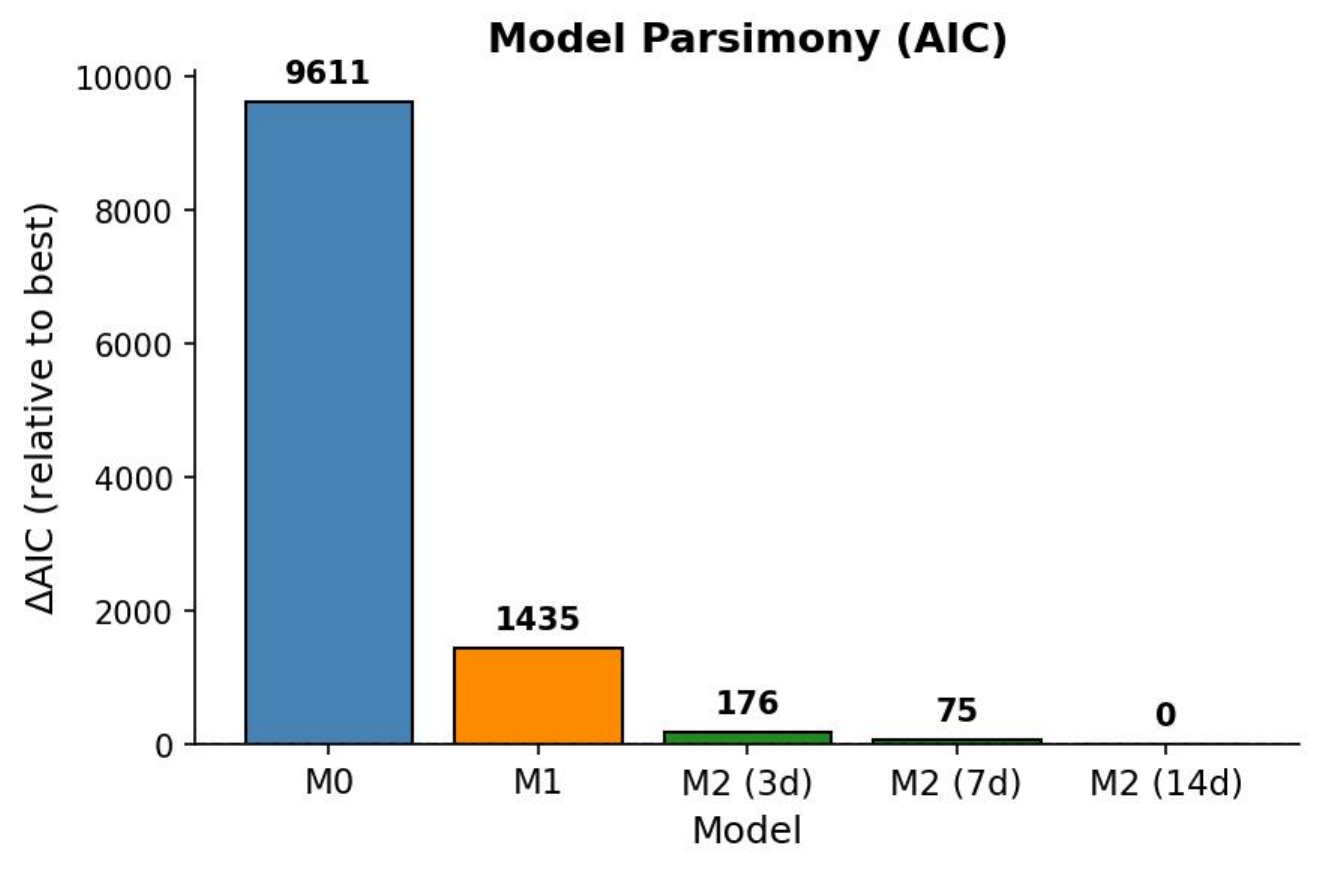}
\end{minipage}
\\
\begin{minipage}[t]{\columnwidth}
(b)\\
\vspace{0.1cm}
\includegraphics[width=\linewidth]{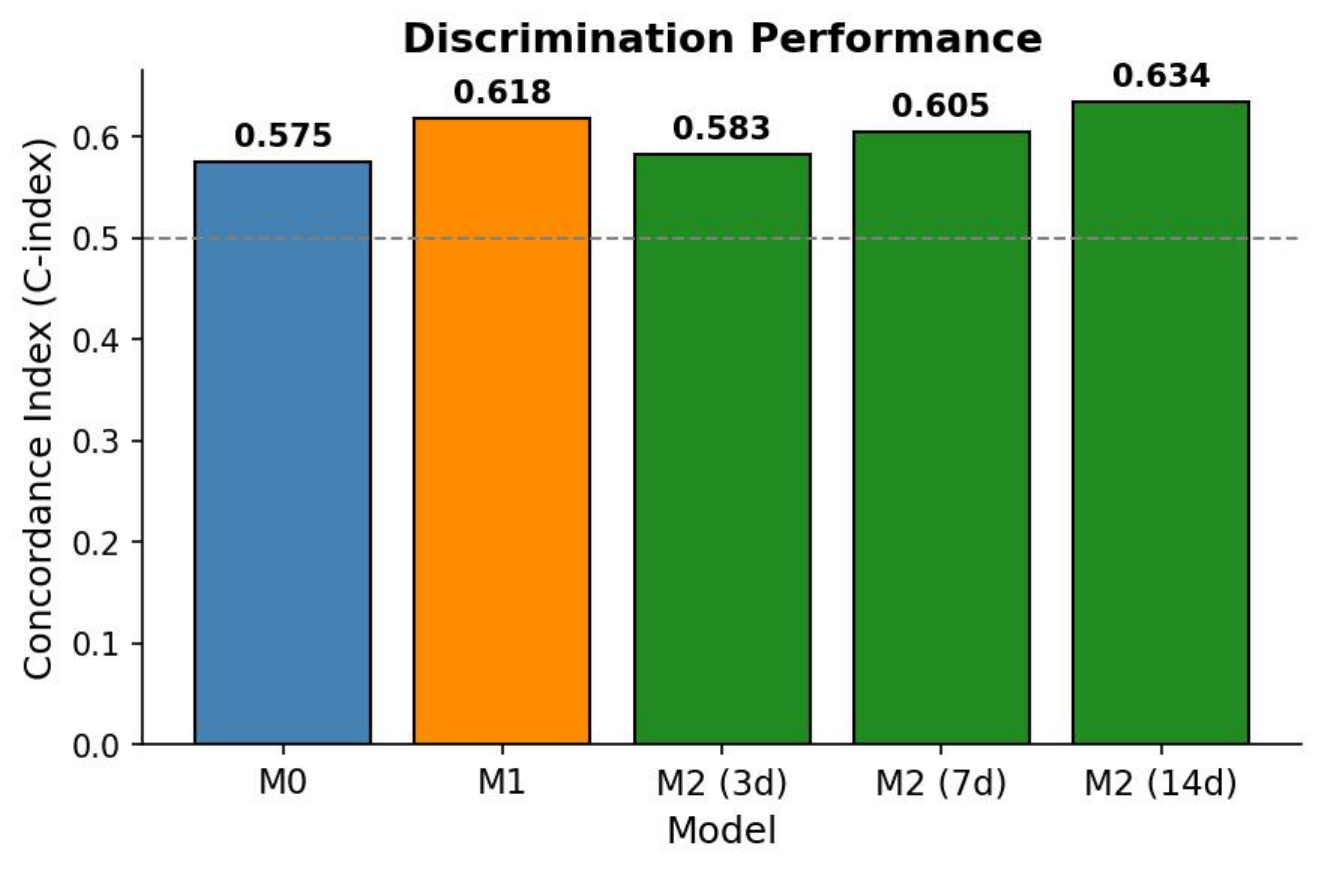}
\end{minipage}
\caption{Performance comparison of the sequential Cox proportional hazards models. (a) Model fit measured by $\Delta$AIC relative to the best-fitting model. (b) Within-sample discrimination measured by Harrell's C-index. Models M0--M2 correspond to the sequential models described in the supplementary statistical analysis, with M2 evaluated using 3-, 7-, and 14-day averaging windows for pre-invasion transmission pressure.}
\label{fig:survival-model-performance}
\vspace{-0.2cm}
\end{figure}

Fig.~\ref{fig:survival-model-performance} shows that adding covariates improved both model fit and discrimination. Relative to M2 with the 14-day pressure window, the $\Delta$AIC values were 9,611 for M0 and 1,435 for M1. Adding pre-invasion transmission pressure to M1 further reduced $\Delta$AIC to 176, 75, and 0 for the 3-, 7-, and 14-day windows, respectively. Thus, the 14-day specification achieved the lowest AIC among the evaluated models.

The C-index showed a similar pattern, increasing from 0.575 for M0 to 0.618 for M1. For M2, the corresponding values were 0.583, 0.605, and 0.634 for the 3-, 7-, and 14-day windows, respectively. These results indicate that adjustment for event-generated epidemic seed size improved within-sample discrimination, while incorporating pre-invasion transmission pressure yielded further improvements. Among the evaluated specifications, M2 with the 14-day window achieved both the lowest AIC and the highest C-index. The agreement between the two metrics supports the conclusion that the longer averaging window provided the strongest statistical fit and discrimination in this analysis.

Despite the improved performance of M2, M1 was retained as the primary adjusted model for the subsequent analyses. This choice reflects the different interpretive roles of the models. M0 estimates the unadjusted association between MGE occurrence and district-level invasion timing, whereas M1 examines how this association changes after adjustment for the size of the event-generated epidemic seed. M2 additionally incorporates transmission pressure measured retrospectively over the period preceding each district's observed arrival time. It therefore characterizes the additional statistical information associated with the recent transmission environment, rather than an initial-state covariate available independently of the observed arrival time. M1 was consequently used for the subsequent parameter-sensitivity and seed-location analyses to maintain a consistent focus on the association between MGEs and invasion timing after adjustment for event-generated seeding. M2 served as a complementary analysis of model performance and the information associated with recent pre-invasion transmission pressure.

Overall, the comparison indicates that event-generated seed size and recent transmission pressure each provide statistical information relevant to district-level invasion timing. However, the association between MGEs and earlier invasion remained substantial after adjustment for seed size in the primary analysis. The additional improvement obtained with M2 suggests that differences in invasion timing are also associated with the transmission environment preceding district-level arrival. These findings characterize relationships within the simulated epidemic trajectories; they do not, on their own, establish causal mechanisms or demonstrate prospective predictive performance.

\section{Sensitivity analysis across transmission conditions}
\label{sec:supplementary-sensitivity-analysis}

% \subsection{Sensitivity to community and event transmission parameters}
To assess whether the association between transmission conditions and district-level epidemic invasion timing was sensitive to epidemiological parameters, we conducted a survival analysis using the AMS\_dance scenario. We considered 25 combinations of the basic reproduction number, $R_0 \in \{1.0,1.5,\ldots,3.0\}$, and the event-specific transmission parameter, $\beta_{\mathrm{event}} \in \{0.1,0.2,\ldots,0.5\}$. Stochastic simulations were repeated for each combination, and district-level epidemic arrival times were extracted.

Each parameter combination was represented by a categorical variable, $C_r$, with $R_0=1.0$ and $\beta_{\mathrm{event}}=0.1$ as the reference condition. This specification allowed each combination to be compared with the reference without imposing linear effects of either parameter. We fitted three Cox proportional hazards models, following the sequential structure of the primary analysis:
\begin{align}
\text{P0:} \quad
h_i(t)
&=
h_0(t) \exp\left( \theta_{\mathrm{param}}(\mathbf{x}_i) \right),
\\
\text{P1:} \quad
h_i(t)
&=
h_0(t)
\exp\left( \theta_{\mathrm{param}}(\mathbf{x}_i) + \theta_Z\log(Z_i+1) \right),
\\
\begin{split}
\text{P2:} \quad
h_i(t)
&=
h_0(t) \\
&
\exp\left(
\theta_{\mathrm{param}}(\mathbf{x}_i)
+ \theta_Z\log(Z_i+1)
+ \theta_{\Lambda}\bar{\Lambda}_i^{(W)} \right).	
\end{split}	
\end{align}
\noindent
Here, $\theta_{\mathrm{param}}(\mathbf{x}_i)$ denotes the coefficient associated with the transmission-parameter combination assigned to district $i$. As in the primary analysis, $Z_i$ represents the event-generated epidemic seed and $\bar{\Lambda}_i^{(W)}$ denotes the mean daily pre-invasion transmission pressure defined in Eq.~\ref{eq:avg-infection-force}. As in the primary analysis, cluster-robust standard errors were calculated using the simulation replicate as the clustering unit. %Models P0--P2 therefore parallel Models M0--M2 and allow us to assess whether the apparent association between transmission conditions and invasion timing changes after accounting for event-generated epidemic seeding and recent transmission pressure.

% Here, $h_{ir}(t)$ denotes the invasion hazard for district $i$ in simulation replicate $r$, and $\theta_{\mathrm{param}}(C_r)$ represents the coefficient associated with the parameter combination assigned to that replicate. The event-generated epidemic seed size, $Z_r$, and mean pre-invasion transmission pressure, $\bar{\Lambda}_{ir}^{(W)}$, are defined in Section~\ref{sec:supplementary-statistical-analysis}. As in the primary analysis, cluster-robust standard errors were calculated using the simulation replicate as the clustering unit.

The main sensitivity analysis focused on P0 and P1. P0 estimates the unadjusted association between the parameter combinations and invasion timing, whereas P1 examines how these associations change after adjustment for event-generated seed size. P2 additionally incorporates pre-invasion transmission pressure. Fig.~\ref{fig:parameter-heatmap} summarizes the hazard ratios from P0 and P1 across the parameter space, together with the change in estimates after adjustment.

\begin{figure*}
\centering
\includegraphics[width=17cm]{figures/parameter_analysis_AMS_dance_seeds0_heatmap.pdf}
\caption{Hazard ratios across epidemiological parameter combinations for the AMS\_dance scenario. (a) Estimates from the parameter-only Cox model (P0). (b) Estimates after adjustment for the logarithm of event-generated epidemic seed size (P1). (c) Difference between the P0 and P1 hazard-ratio estimates. All estimates are relative to the reference condition ($R_0=1.0$, $\beta_{\mathrm{event}}=0.1$).}
\label{fig:parameter-heatmap}
\end{figure*}

Before adjustment for seed size, the estimated hazard ratios generally increased with both $R_0$ and $\beta_{\mathrm{event}}$, indicating associations between more transmissible conditions and earlier district-level invasion. At $\beta_{\mathrm{event}}=0.5$, the hazard ratios were 1.40, 1.75, 2.06, and 2.32 for $R_0=1.5$, 2.0, 2.5, and 3.0, respectively, relative to the reference condition.

After adjustment for event-generated seed size, the association with $\beta_{\mathrm{event}}$ was substantially attenuated, particularly at moderate values of $R_0$. For example, at $R_0=1.5$, the adjusted hazard ratios ranged from 0.98 to 1.07 across the five values of $\beta_{\mathrm{event}}$, compared with 0.82 to 1.40 in P0. In contrast, the association with community transmissibility remained pronounced. At $\beta_{\mathrm{event}}=0.1$, the adjusted hazard ratio increased from 0.98 at $R_0=1.5$ to 1.34, 1.66, and 1.99 at $R_0=2.0$, 2.5, and 3.0, respectively.

These results indicate that the association between event-specific transmission and invasion timing was substantially attenuated after accounting for event-generated seed size, whereas the association with community transmissibility remained pronounced. This pattern is consistent with seed size accounting statistically for part of the association between event transmission conditions and subsequent invasion timing. The persistence of the $R_0$ association suggests that community transmissibility remains associated with the subsequent spatial propagation of infection, beyond differences in the initial event-generated seed. These interpretations concern associations within the simulated system and do not independently establish causal pathways.

%\subsection{Sensitivity to the location of the initial epidemic seed}
The descriptive analysis of arrival-time shifts revealed heterogeneity across seed--destination pairs, see Fig.~\ref{fig:arrival_matrix_diff}, suggesting that the spatial origin of the epidemic may affect the magnitude of MGE-associated changes in invasion timing. We therefore repeated the primary Cox analysis for three seed districts representing contrasting spatial and network characteristics: District 0, a centrally located district with a medium-sized population; District 7, a peripheral district with a large population; and District 18, a peripheral district with high network connectivity. The analysis used the representative parameter setting $R_0=1.5$, $\beta_{\mathrm{event}}=0.5$, and $I_{\mathrm{ss}}=1$, with the same MGE scenarios and no-event baseline as in the primary analysis.

\begin{table*}
\normalsize
\setlength{\tabcolsep}{8pt}
\renewcommand{\arraystretch}{1.2}
\centering
\caption{Sequential Cox model estimates for three initial epidemic seed locations under $R_0=1.5$, $\beta_{\mathrm{event}}=0.5$, and $I_{\mathrm{ss}}=1$. M0 includes the MGE indicator only; M1 additionally adjusts for the logarithm of event-generated epidemic seed size. Values are hazard ratios (HRs) with 95\% confidence intervals.}
\label{table:seed-location-robustness}
\resizebox{11.0cm}{!}{
\begin{tabular}{cccc}
	\hline\hline
	Seed location & M0 HR & M1 HR & M1 seed-size HR \\
	\hline
	District 0 & 2.90 (2.70--3.11) & 2.66 (2.47--2.86) & 1.45 (1.43--1.47) \\
	District 7 & 1.99 (1.85--2.14) & 1.74 (1.61--1.87) & 1.83 (1.80--1.86) \\
	District 18 & 2.20 (2.05--2.36) & 1.95 (1.81--2.10) & 1.72 (1.70--1.75) \\
	\hline\hline
\end{tabular}
}
\vspace{+0.3cm}
\end{table*}

\begin{figure*}
\centering
\includegraphics[width=17cm]{figures/seed_location_robustness_selected.pdf}
\vspace{+0.3cm}
\caption{Sensitivity of the Cox model estimates to the initial epidemic seed location. Results are shown for three seed districts with contrasting spatial and network characteristics. (a) MGE hazard ratios from M0 and M1, with 95\% confidence intervals; the dashed line indicates HR = 1. (b) Harrell's C-index for M0 and M1. (c) Improvement in AIC-based fit after adjustment for seed size, calculated as $\mathrm{AIC}_{M0}-\mathrm{AIC}_{M1}$; positive values indicate a lower AIC for M1.}
\label{fig:seed-location-robustness}
\end{figure*}

The association between MGEs and earlier district-level invasion remained positive across all three seed locations, see Table~\ref{table:seed-location-robustness} and Fig.~\ref{fig:seed-location-robustness}(a). In M0, the estimated HR was largest for District 0 (2.90, 95\% CI: 2.70--3.11), followed by District 18 (2.20, 95\% CI: 2.05--2.36) and District 7 (1.99, 95\% CI: 1.85--2.14). After adjustment for event-generated seed size, the corresponding MGE HRs decreased to 2.66 (95\% CI: 2.47--2.86), 1.95 (95\% CI: 1.81--2.10), and 1.74 (95\% CI: 1.61--1.87), respectively. Thus, although the magnitude varied with seed location, the estimated association remained positive after adjustment in all three cases.

Event-generated seed size was also positively associated with earlier invasion across the three locations. Its estimated HR was 1.45 (95\% CI: 1.43--1.47) for District 0, 1.83 (95\% CI: 1.80--1.86) for District 7, and 1.72 (95\% CI: 1.70--1.75) for District 18. These estimates indicate that the association between seed size and invasion timing also varied across the selected spatial configurations.

Model performance improved after adjustment for seed size at all three locations, see Fig.~\ref{fig:seed-location-robustness}(b, c). The C-index increased from 0.575 to 0.618 for District 0, from 0.553 to 0.627 for District 7, and from 0.559 to 0.622 for District 18. The corresponding reductions in AIC were 8,176, 18,514, and 15,258. Thus, M1 provided better within-sample discrimination and AIC-based fit than M0 for each of the three seed locations.

Taken together, the seed-location analysis indicates that the positive association between MGEs and earlier invasion, including its persistence after adjustment for event-generated seed size, was not restricted to the default seed location. However, the magnitude of the estimated associations varied across the three selected districts. Because only three locations were examined, these results demonstrate robustness across the selected configurations rather than across all possible epidemic origins.

%\subsection{Synthesis}
The two sensitivity analyses provide complementary evidence about variation in district-level invasion timing. Across the tested transmission conditions, adjustment for event-generated seed size substantially attenuated the association with $\beta_{\mathrm{event}}$, whereas the association with $R_0$ remained pronounced. Across the three selected seed locations, the association between MGEs and earlier invasion remained positive after adjustment for seed size, although its magnitude varied by location.

These results are consistent with event-generated seeding accounting statistically for part of the association between MGEs and earlier spatial invasion, while community transmissibility and the spatial context of the epidemic remain relevant to subsequent propagation. The findings support the robustness of the principal statistical patterns across the parameter combinations and seed locations examined, without establishing causal mediation or implying that the estimates generalize to every epidemiological condition or epidemic origin.

% bibliography
%\bibliographystyle{IEEEtran}
%\bibliography{refs_20260828a_sorted}

\vfill
\clearpage

\end{document}